\documentclass[reprint,amsmath,amssymb,superscriptaddress,prb]{revtex4-2}

\usepackage{graphicx}
\usepackage{dcolumn}
\usepackage{bm}
\usepackage{color}
\usepackage{hyperref}
\usepackage{subcaption}
\usepackage{gensymb}
\usepackage{csquotes}

\begin{document}

\title{Quantum effects of Coulomb explosion simulations revealed by time-dependent density-functional theory}

\author{Samuel S. Taylor}
\affiliation{Department of Physics and Astronomy, Vanderbilt University, Nashville, Tennessee, 37235, USA}

\author{Cody Covington}
\affiliation{Department of Chemistry, Austin Peay State University,
Clarksville, USA}

\author{K\'alm\'an Varga}
\email{kalman.varga@vanderbilt.edu}
\affiliation{Department of Physics and Astronomy, Vanderbilt University, Nashville, Tennessee, 37235, USA}

\begin{abstract}
This study investigates the influence of quantum effects on Coulomb explosion dynamics using time-dependent density functional theory (TDDFT) simulations, comparing classical, semi-classical, and quantum approaches. The goal is to elucidate how electron dynamics affect the kinetic energy, angular distribution, and final velocities of ejected ions. The results indicate that quantum effects result in lower kinetic energies all ions, deviating from classical predictions. Furthermore, quantum simulations exhibit broader angular distributions and more diverse ion trajectories, aligning closely with experimental observations. The research also highlights the role of laser intensity and the resultant ionization in enhancing quantum effects, particularly in modifying ion velocities and distributions. These findings provide a deeper understanding of the role of electron dynamics in Coulomb explosions, offering valuable insights for both experimental and theoretical studies of molecular fragmentation.
\end{abstract}

\maketitle

\section{Introduction}

Recent advancements in ultrafast, high-intensity laser technology have enabled real-time studies of electron and ion dynamics in extreme environments~\cite{Keller2003,RevModPhys.81.163,https://doi.org/10.1002/wcms.1430}. This progress has facilitated extensive investigations into Coulomb explosion—a phenomenon in which a laser ionizes a molecule, causing the remaining positively charged ions to repel each other. The Coulomb explosion process not only provides insights into the behavior of atoms in these high-energy environments~\cite{https://doi.org/10.1002/wcms.1430}, but also has diverse applications, including molecular fragmentation~\cite{D3CP01740K,Hishikawa1998,https://doi.org/10.1002/cphc.201501118,Luzon2019,Cornaggia_1992,C9RA02003A,XU2009255,XU2010119,10.1063/1.3561311,Severt2024,Kwon2023,10.1063/1.5070067,Bhattacharyya2022,PhysRevA.96.043415}, generation of bright keV x-ray photons~\cite{PhysRevLett.75.3122,McPherson1994}, production of highly energetic electrons~\cite{PhysRevLett.77.3343}, structural dynamics~\cite{10.1063/1.5041381,doi:10.1126/science.abc2960,Erattupuzha_2017,Légaré_2006,PhysRevLett.74.3780,PhysRevA.101.012707,PhysRevA.109.023112,doi:10.1126/science.adk1833,PhysRevLett.131.143001,D3CP01740K,Ekanayake2017,Ekanayake2018}, and molecular imaging~\cite{doi:10.1126/science.244.4903.426,annurev:/content/journals/10.1146/annurev-physchem-090419-053627,doi:10.1126/science.1246549,Mogyorosi2020,doi:10.1126/science.1240362,10.1063/5.0024833,YATSUHASHI201852,C1CP21345H,D2CP01114J,Crane2021,PhysRevLett.99.258302,PhysRevA.99.023423,Corrales2019,10.1063/5.0098531,Zhang2022,10.1063/5.0200389,PhysRevA.107.023104,doi:10.1080/23746149.2022.2132182,PhysRevLett.130.093001,Howard2023,PhysRevLett.107.063201,PhysRevResearch.4.013029,Unwin2023,Boll2022,lam2024imagingcoupledvibrationalrotational,C7CP01379E,PhysRevLett.132.123201,10.1063/1.4982220,PhysRevA.71.013415,C0CP02333G,PhysRevA.91.053424,Schouder_2021,PhysRevA.103.042813}.

Experimental studies of the Coulomb explosion process frequently rely on computational approaches as theoretical models to complement and support experimental observations. In these studies, Coulomb explosion simulations are often used to compare the theoretical predictions with experimental data, particularly regarding the kinetic energy and angular distribution of the ejected ions. However, different simulation methods can yield varying results. Most experimental work employs a classical Coulomb explosion model, in which ions are treated as point charges that repel each other solely via Coulomb interactions, without accounting for the presence of physical electrons~\cite{PhysRevResearch.4.013029,Unwin2023,Boll2022,PhysRevLett.132.123201,lam2024imagingcoupledvibrationalrotational,10.1063/1.4982220,C7CP01379E,Bhattacharyya2022,PhysRevA.71.013415,C0CP02333G,PhysRevA.91.053424,PhysRevA.96.043415,Schouder_2021,PhysRevA.103.042813,andré2024partialorientationretrievalproteins,andré2024proteinstructureclassificationbased}. This approach neglects quantum effects. In real physical systems, electron interactions are present and influence molecular dynamics, even without a laser field. When the intense electric field of the laser is introduced, these electron effects are amplified, further shaping the molecular dynamics both during and after the Coulomb explosion process.

A recent study employs trajectory surface hopping (TSH) to investigate the quantum effects of Coulomb explosion~\cite{PhysRevA.109.052813}. While the TSH approach used effectively captures nonadiabatic nuclear dynamics through stochastic transitions between electronic states, it often approximates electron density dynamics, treating them as either largely static or discretely changing~\cite{https://doi.org/10.1002/wcms.1158,Barbatti2022}. In contrast, time-dependent density functional theory (TDDFT) models electron dynamics by continuously evolving the electron density in real time across all states. This continuous evolution provides a more comprehensive understanding of the ionization processes central to Coulomb explosions, as TDDFT resolves the gradual depletion of electron density and the concurrent accumulation of positive charge within the system. Furthermore, the explicit propagation of the wavefunction in TDDFT accounts for complex electron correlation effects, enabling it to capture detailed electron dynamics such as electron localization and density flow. These effects can significantly influence the energetic pathways of fragmenting ions. TDDFT stands out as a particularly effective method for studying the quantum effects of Coulomb explosion, as it accurately models the interactions between electrons and ions. By simulating both the electrons and molecular orbitals, TDDFT captures the quantum effects that have important implications for experimental studies. As a result, TDDFT has demonstrated strong agreement with Coulomb explosion experiments~\cite{PhysRevA.89.023429,
PhysRevA.91.023422,PhysRevA.92.053413,PhysRevA.95.052701,PhysRevA.86.043407}.

In this study, we use TDDFT-based quantum Coulomb explosion simulations to compare with classical simulations and semi-classical simulations, the latter being a hybrid of the two previous methods. The motivation for this work stems from a recent study that employ classical Coulomb explosion models to generate Newton plots by mapping the final momentum of each ion, thus enabling the determination of molecular structure~\cite{PhysRevLett.132.123201}. These Newton-plots can be compared to the experimentally measured 
three dimensional ion momenta distributions.   The experimental ion momenta patterns appear significantly broader than the distributions obtained by classical simulations \cite{PhysRevLett.132.123201}. In this work, we will show that the broadening is due to quantum effects.

Isoxazole, used in the recent study\cite{PhysRevLett.132.123201}, will also serve as a comparison molecule in this study to assess the impact of quantum effects on Coulomb explosion imaging and whether the molecular structure can be reconstructed with the same clarity and accuracy. We will investigate the ion kinematics, kinetic energy, and angular distributions throughout the Coulomb explosion process for acetylene, butane, and isoxazole.

\section{Computational Method}
The quantum simulations were performed using TDDFT for modeling the electron dynamics on a 
real-space grid with real-time propagation \cite{Varga_Driscoll_2011a}, 
with the Kohn-Sham (KS) Hamiltonian of the following form
\begin{equation}
\begin{split}
\hat{H}_{\text{KS}}(t) = -\frac{\hbar^2}{2m} \nabla^2 + V_{\text{ion}}(\mathbf{r},t) 
+ V_{\text{H}}[\rho](\mathbf{r},t) \\
+ V_{\text{XC}}[\rho](\mathbf{r},t) + V_{\text{laser}}(\mathbf{r},t).
\end{split}
\label{eq:hamiltonian}
\end{equation}
Here, $\rho$ is the electron density which is defined as the density sum over all occupied orbitals:
\begin{equation}
\rho(\mathbf{r},t) = \sum_{k=1}^{N_{\text{orbitals}}} 2|\psi_k(\mathbf{r},t)|^2,
\end{equation}
where the coefficient 2 accounts for there being two electrons in each orbital (via spin degeneracy) and $k$ is a quantum number labeling each orbital. 

$V_{ion}$ in eq.~\ref{eq:hamiltonian} is the external potential due to the ions, represented by employing norm-conserving pseudopotentials centered at each ion as given by Troullier and Martins~\cite{PhysRevB.43.1993}. $V_{H}$ is the Hartree potential, defined as
\begin{equation}
V_H(\mathbf{r}, t) = \int \frac{\rho(\mathbf{r}', t)}{|\mathbf{r} - \mathbf{r}'|} \, d\mathbf{r}',
\end{equation}
and accounts for the electrostatic Coulomb interactions between electrons. The term $V_{XC}$ is the exchange-correlation potential, which is approximated by the adiabatic local-density approximation (ALDA), obtained from a parameterization to a homogeneous electron gas by Perdew and Zunger~\cite{PhysRevB.23.5048}. The last term in eq.~\ref{eq:hamiltonian}, $V_{laser}$ is the  time–dependent potential due to the electric field of the laser, and is described using the dipole approximation, $V_{\text{laser}} = \mathbf{r} \cdot \mathbf{E}_{\text{laser}}(t)$. The electric field $\mathbf{E}_{\text{laser}}(t)$ is given by
\begin{equation}
\mathbf{E}_{\text{laser}}(t) = E_{\text{max}} \exp\left[-\frac{(t - t_0)^2}{2a^2}\right] \sin(\omega t) \mathbf{\hat{k}},
\label{eq:laser}
\end{equation}
where the parameters $E_{\text{max}}$, $t_{0}$, and $a$ define the maximum amplitude, initial position of the center, and the width of the Gaussian envelope, respectively. $\omega$ describes the frequency of the laser and $\mathbf{\hat{k}}$ is the unit vector in the polarization direction of the electric field. 

At the beginning of the TDDFT calculations, the ground state of the system is prepared by performing a Density-Functional Theory (DFT) calculation. With these initial conditions in place, we then proceed to propagate the Kohn–Sham orbitals, $\psi_{k}(\mathbf{r},t)$ over time by using the time-dependent KS equation, given as 
\begin{equation}
i \frac{\partial \psi_k(\mathbf{r}, t)}{\partial t} = \hat{H} \psi_k(\mathbf{r}, t).
\label{eq:tdks}
\end{equation}
Eq.~\ref{eq:tdks} was solved using the following time propagator
\begin{equation}
\psi_k(\mathbf{r}, t + \delta t) = \exp\left(-\frac{i \hat{H}_{\text{KS}}(t) \delta t}{\hbar}\right) \psi_k(\mathbf{r}, t).
\end{equation}
This operator is approximated using a fourth-degree Taylor expansion, given as
\begin{equation}
\psi_k(\mathbf{r}, t + \delta t) \approx \sum_{n=0}^{4} \frac{1}{n!} \left(\frac{-i \delta t}{\hbar} \hat{H}_{\text{KS}}(t)\right)^n \psi_k(\mathbf{r}, t).
\end{equation}
The operator is applied for $N$ time steps until the final time, $t_{final} = N \cdot \delta t$, is obtained. A time step of $\delta t = 1$ as was used in the simulations.

In real-space TDDFT, the Kohn-Sham orbitals are represented at discrete points in real space. These points are organized on a uniform rectangular grid. The accuracy of the simulations is determined by the grid spacing, which is the key parameter that can be adjusted. In our simulations, we used a grid spacing of 0.3 Å and placed 100 points along each of the $x$, $y$, and $z$ axes.

To enforce boundary conditions, we set the Kohn-Sham orbitals to zero at the edges of the simulation cell. However, when a strong laser field is applied, ionization can occur, potentially causing unphysical reflections of the wavefunction at the cell boundaries. To address this issue, we implemented a complex absorbing potential (CAP) to dampen the wavefunction as it reaches the boundaries. The specific form of the CAP used in our simulations, as described by Manopolous~\cite{10.1063/1.1517042}, is given by:
\begin{equation}
- i w(x) = -i \frac{\hbar^2}{2m} \left(\frac{2\pi}{\Delta x}\right)^2 f(y),
\end{equation}
where $x_{1}$ is the start and $x_{2}$ is the end of the absorbing region, $\Delta x = x_{2} - x_{1}$, $c = 2.62$ is a numerical constant, $m$ is the electron’s mass, and
\begin{equation}
f(y) = \frac{4}{c^2} \left( \frac{1}{(1 + y)^2} + \frac{1}{(1 - y)^2} - 2 \right), \quad y = \frac{x - x_1}{\Delta x}.
\end{equation}

As the molecule is ionized by the laser field, the electron density is directed towards the CAP. Additionally, the ejected fragments move towards the CAP, carrying their electron density. When any electron density into contact with the CAP, it is absorbed. Consequently, the total electron number
\begin{equation}
N(t) = \int_V \rho(\mathbf{r}, t) \, d^3x,
\end{equation}
where $V$ is the volume of the simulation box, will diverge from the initial electron number, $N(0)$. We interpret $N(0) - N(t)$ as the total number of electrons ejected from the simulation box.

Motion of the ions in the simulations were treated classically. Using the Ehrenfest theorem
, the quantum forces on the ions due to the electrons are given by the derivatives 
of the expectation value of the total electronic energy with respect to the ionic positions. 
These forces are then fed into Newton’s Second Law, giving
\begin{equation}
\begin{split}
M_i \frac{d^2 \mathbf{R}_i}{dt^2} = Z_i \mathbf{E}_{\text{laser}}(t) 
+ \sum_{j \neq i}^{N_{\text{ions}}} \frac{Z_i Z_j (\mathbf{R}_i - \mathbf{R}_j)}{|\mathbf{R}_i - \mathbf{R}_j|^3}\\ 
- \nabla_{\mathbf{R}_i} \int V_{\text{ion}}(\mathbf{r}, \mathbf{R}_i) \rho(\mathbf{r}, t) \, d\mathbf{r},
\end{split}
\label{eq:newton-law-quantum}
\end{equation}
where $M_{i}$, $Z_{i}$, and $\mathbf{R}_{i}$ are the mass, pseudocharge (valence), and position of the $i$-th ion, respectively, and $N_{\text{ions}}$ is the total number of ions. This differential equation was time propagated using the Verlet algorithm at every time step $\delta t$. This approach has been successfully used to describe the Coulomb explosion of molecules~\cite{PhysRevA.89.023429,
PhysRevA.91.023422,PhysRevA.92.053413,PhysRevA.95.052701,PhysRevA.86.043407}.

In each quantum simulation, the ion velocities are initialized using a Boltzmann distribution corresponding to 300 K. This random initialization facilitates the exploration of various fragmentation pathways during the Coulomb explosion. If the ion velocities were held constant, the simulations would yield identical dissociation outcomes, thus constraining the statistical variability of the results.

The classical Coulomb explosion simulations were designed to closely mirror the quantum simulations for ease of comparison. In the classical approach, each atom’s charge was assumed to remain constant throughout the simulation. The specific charge assigned to each atom was based on the average charge observed post-ionization from the laser in the quantum Coulomb explosion simulations. This method ensured consistent charge states across both the classical and quantum cases, aligning with the typical assumption in classical simulations of fixed charge states. Here, the only variable between different classical simulations is the seed used in the Boltzmann distribution calculation.

With no external laser field applied, the sole force acting on each ion in the classical simulations is the Coulomb repulsion, which we model using Newton's Second Law:
\begin{equation}
M_i \frac{d^2 \mathbf{R}_i}{dt^2} =  \sum_{j \neq i}^{N_{\text{ions}}} \frac{Z_i Z_j (\mathbf{R}_i - \mathbf{R}_j)}{|\mathbf{R}_i - \mathbf{R}_j|^3}\\ ,
\end{equation}
where $M_{i}$, $Z_{i}$, and $\mathbf{R}_{i}$ represent the mass, charge, and position of the $i$-th ion, respectively, and $N_{\text{ions}}$ is the total number of ions (consistent with Eq.~\ref{eq:newton-law-quantum}). As with the quantum simulations, initial ion velocities were assigned using a Boltzmann distribution at 300 K. The simulation then evolved over time using the Verlet algorithm at each time step $\delta t = 1$ as.

The semi-classical simulation method is a hybrid of both classical and quantum approaches. In this method, the simulation begins with a quantum treatment of the molecule during the laser-induced ionization phase. At a chosen time, $t^{*}$, once ionization has finished, the simulation transitions to a classical approach. At this transition point, $t^{*}$, each atom’s position, velocity, and charge are taken from the quantum simulation values at $t^{*}$, effectively continuing the simulation with the classical approach from that point forward. This approach will elucidate the quantum effects between the ion's electrons after the laser-induced ionization.

In all simulations, each molecule was positioned at the center of the simulation box, with its longest molecular axis aligned along the x-axis and its shortest axis oriented along the z-axis. For every quantum (and thus also every semi-classical) simulation, the electric field was polarized along the x-axis to maximize ionization~\cite{PhysRevA.92.053413}. In experimental settings, this precise molecular alignment can be achieved depending on the alignment scheme and the molecule's polarizability \cite{annurev:/content/journals/10.1146/annurev-physchem-090419-053627}.

For each simulation, the laser parameters were held constant: the wavelength was set to 800 nm, the pulse duration to 7 fs (FWHM), and the peak electric field to 14 V/\AA. These parameters were motivated by the recent study~\cite{PhysRevLett.132.123201}, with a reduced duration to accommodate the simulation window. In addition, a separate set of simulations for butane was performed with double the electric field strength (28 V/\AA).

In the following section, we present the results of the classical, semi-classical, and quantum Coulomb explosion simulations for acetylene (C\textsubscript{2}H\textsubscript{2}), butane (C\textsubscript{4}H\textsubscript{10}), and isoxazole (C\textsubscript{3}H\textsubscript{3}NO) based on a large number of simulations conducted for each molecule. Given that classical Coulomb models do not explicitly account for electrons, the results focus on comparing the properties of the ions—specifically their positions, velocities, and angular distributions—throughout the Coulomb explosion processes.

\section{Results}

\begin{figure}[ht!]
    \centering
    \includegraphics[width=0.95\columnwidth]{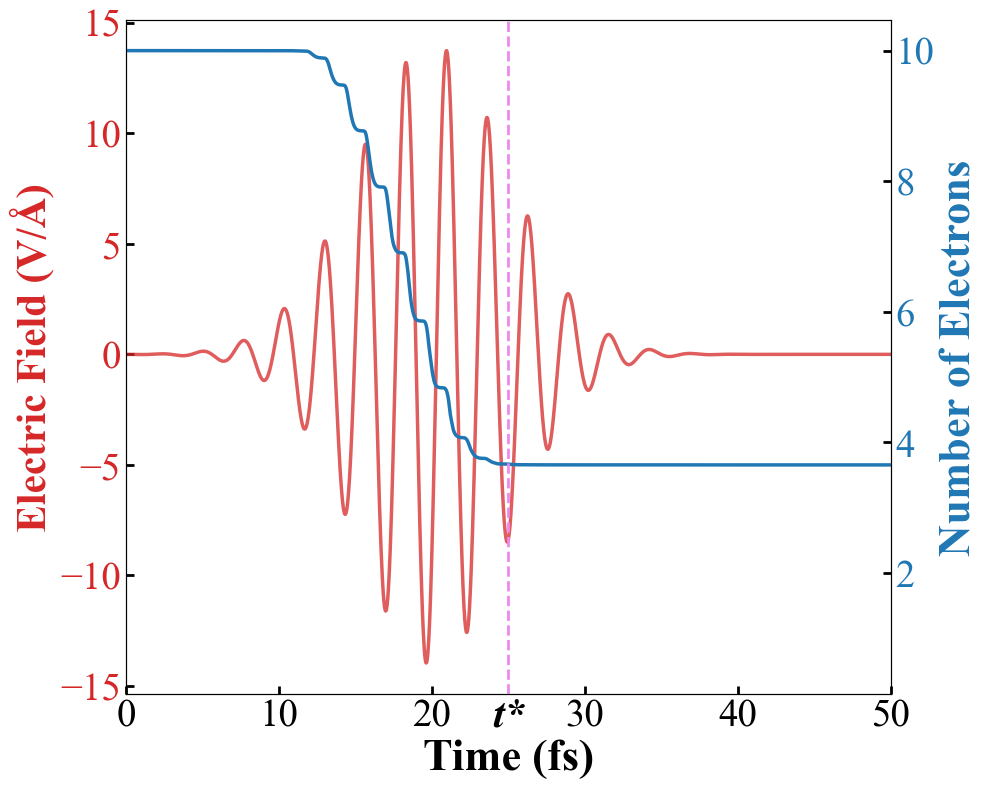}
    \caption{Laser pulse intensity profile and electron count as functions of time in a TDDFT C\textsubscript{2}H\textsubscript{2} Coulomb explosion simulation. The pink dashed line at $t^{*} = 25$ fs marks the transition point from the ionization phase to the classical phase in the semi-classical simulation. Before $t^{*}$, the quantum description governs the ionization process, while after $t^{*}$, the simulation continues with a classical treatment of the system.}
    \label{fig:pulse-electron-c2h2}
\end{figure}

\subsection{Acetylene (C\textsubscript{2}H\textsubscript{2}) Dynamics}

The laser pulse applied in the TDDFT quantum simulations of C\textsubscript{2}H\textsubscript{2} Coulomb explosion is shown in Fig.~\ref{fig:pulse-electron-c2h2}. In the semi-classical approach, the transition from quantum to classical dynamics occurs at \( t^* = 25 \, \text{fs} \), chosen as the time point after which the laser ceases to ionize the molecule. This is evident in Fig.~\ref{fig:pulse-electron-c2h2}, where the electron count stabilizes at 25 fs. The electron count displayed throughout the simulations is derived from a single quantum TDDFT simulation. In the classical Coulomb explosion model for C\textsubscript{2}H\textsubscript{2}, the carbon atoms are assigned charges of $2.07^{+}$ and $2.26^{+}$, respectively, while each hydrogen atom is assigned a charge of $1^{+}$. These values represent average charges calculated over 200 TDDFT simulations. In all of the C\textsubscript{2}H\textsubscript{2} simulations, the CAP was set at -8 and 8 \AA\ on each axis. 

\begin{figure*}[ht!]
    \centering
    \begin{subfigure}[b]{0.45\textwidth} 
        \centering
        \includegraphics[width=\textwidth]{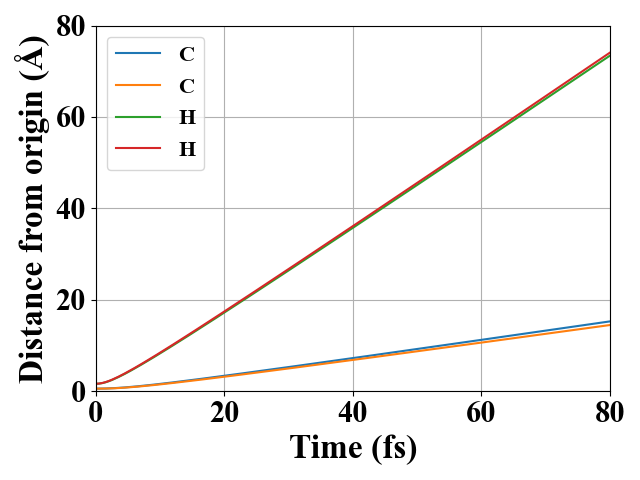}
        \caption{Classical method ion distances}
        \label{fig:c2h2-classical-positions}
    \end{subfigure}
    \hfill
    \begin{subfigure}[b]{0.45\textwidth} 
        \centering
        \includegraphics[width=\textwidth]{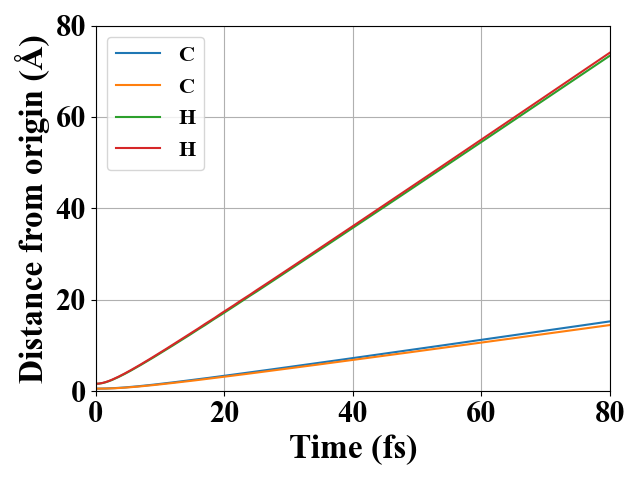}
        \caption{Classical method (with pulse) ion distances}
        \label{fig:c2h2-classical-pulse-positions}
    \end{subfigure}
    
    \vspace{0.3cm} 
    \begin{subfigure}[b]{0.45\textwidth}
        \centering
        \includegraphics[width=\textwidth]{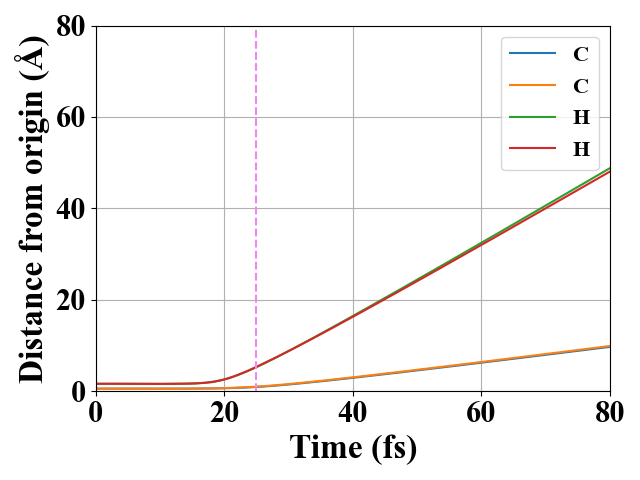}
        \caption{Semi-classical method ion distances}
        \label{fig:c2h2-semi-classical-positions}
    \end{subfigure}
    \hfill
    \begin{subfigure}[b]{0.45\textwidth}
        \centering
        \includegraphics[width=\textwidth]{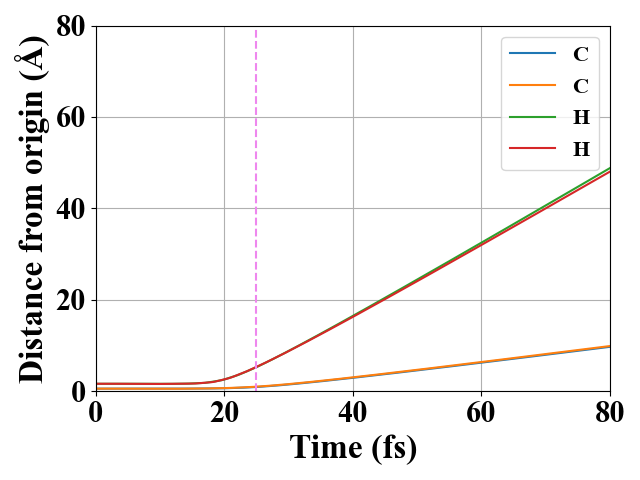}
        \caption{Semi-classical method (with pulse) ion distances}
        \label{fig:c2h2-semi-classical-pulse-positions}
    \end{subfigure}

    \vspace{0.3cm} 
    \begin{subfigure}[b]{0.45\textwidth}
        \centering
        \includegraphics[width=\textwidth]{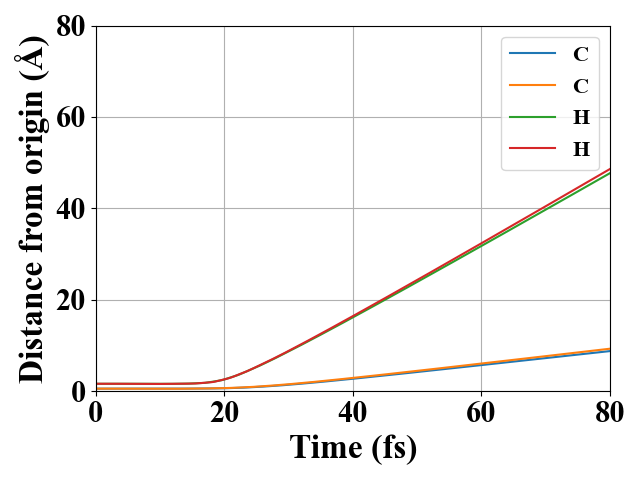}
        \caption{Quantum method ion distances}
        \label{fig:c2h2-quantum-positions}
    \end{subfigure}
    \caption{Comparison of distances across classical, classical (with the pulse included), quantum, semi-classical, and semi-classical (with the pulse included) models of C\textsubscript{2}H\textsubscript{2} Coulomb explosion with 14 V/\AA. The pink, dashed, vertical line in the semi-classical plots indicates the point of time in the simulation where the method switched from TDDFT to classical.}
    \label{fig:c2h2-position-graphs}
\end{figure*}

\begin{figure*}[ht!]
    \centering
    \begin{subfigure}[b]{0.45\textwidth}
        \centering
        \includegraphics[width=\textwidth]{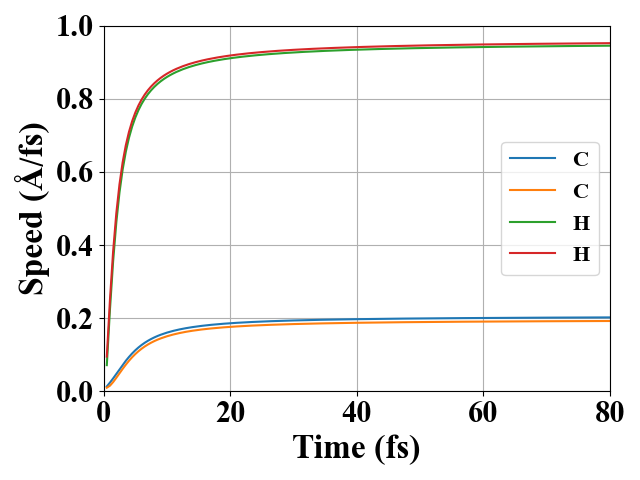}
        \caption{Classical method ion speeds}
        \label{fig:c2h2-classical-velocity}
    \end{subfigure}
    \hfill
    \begin{subfigure}[b]{0.45\textwidth}
        \centering
        \includegraphics[width=\textwidth]{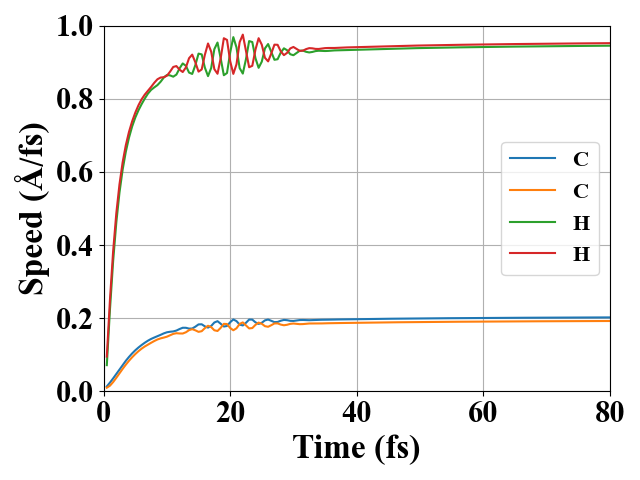}
        \caption{Classical (with pulse) method ion speeds}
        \label{fig:c2h2-classical-pulse-velocity}
    \end{subfigure}
    
    \vspace{0.5cm}
    \begin{subfigure}[b]{0.45\textwidth}
        \centering
        \includegraphics[width=\textwidth]{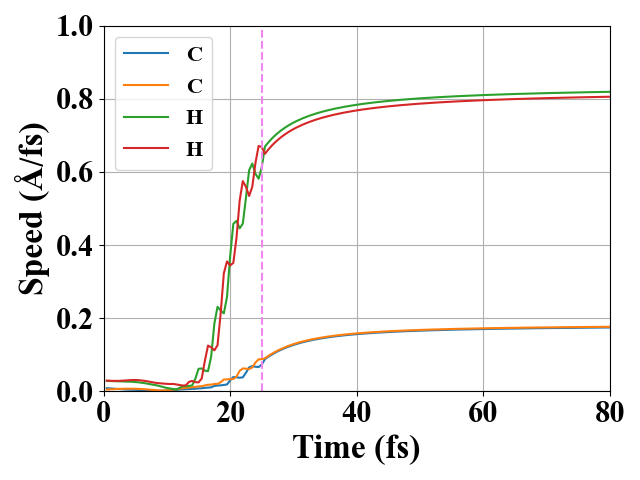}
        \caption{Semi-classical method ion speeds}
        \label{fig:c2h2-semi-classical-velocity}
    \end{subfigure}
    \hfill
    \begin{subfigure}[b]{0.45\textwidth}
        \centering
        \includegraphics[width=\textwidth]{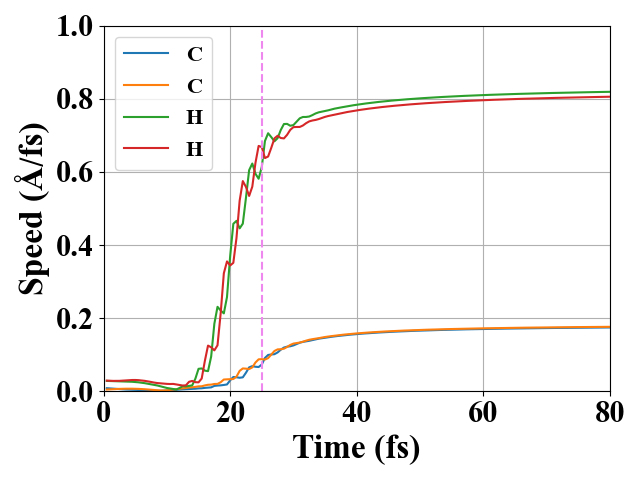}
        \caption{Semi-classical (with pulse) method ion speeds}
        \label{fig:c2h2-semi-classical-pulse-velocity}
    \end{subfigure}

    \vspace{0.5cm}
    \begin{subfigure}[b]{0.45\textwidth}
        \centering
        \includegraphics[width=\textwidth]{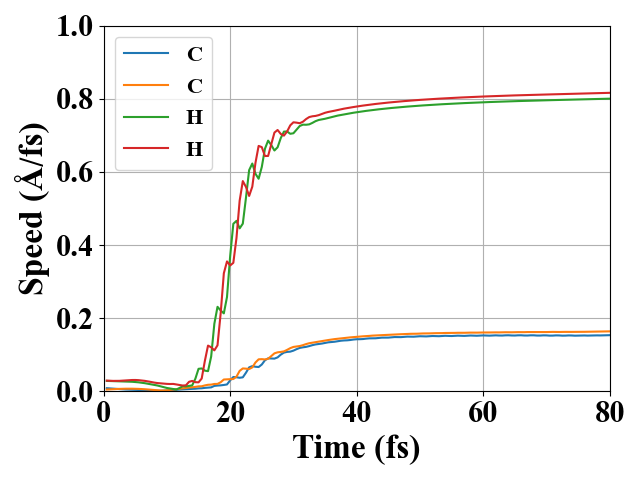}
        \caption{Quantum method ion speeds}
        \label{fig:c2h2-quantum-velocity}
    \end{subfigure}
    \caption{Comparison of speed across classical, classical (with the pulse included), quantum, semi-classical, and semi-classical (with pulse) models of C\textsubscript{2}H\textsubscript{2} Coulomb explosion. The pink, dashed, vertical line in the semi-classical plots indicates the point of time in the simulation where the method switched from TDDFT to classical.}
    \label{fig:c2h2-velocity-graphs}
\end{figure*}

The distances and speeds of ions dissociated from a C\textsubscript{2}H\textsubscript{2} Coulomb explosion are shown in Fig.~\ref{fig:c2h2-position-graphs} and Fig.~\ref{fig:c2h2-velocity-graphs}, respectively. These plots are generated from a single simulation of the acetylene Coulomb explosion, consistent with the electron data used in Fig.~\ref{fig:pulse-electron-c2h2}. The purpose of these figures is to illustrate differences in ionic motion across the three models. As expected, the classical Coulomb model does not capture the effects of the electric field during laser interaction, as it models ionic behavior only after ionization is complete (see Fig.~\ref{fig:c2h2-classical-positions} and Fig.~\ref{fig:c2h2-classical-velocity}). This results in smoother curves: atoms repel each other in near-linear distance and logarithmic speed functions due solely to Coulombic forces.

Including the laser pulse does not affect the final positions (Fig.~\ref{fig:c2h2-classical-pulse-positions}) or velocities (Fig.~\ref{fig:c2h2-classical-pulse-velocity}) compared to simulations without the pulse (Fig.~\ref{fig:c2h2-classical-positions} and Fig.~\ref{fig:c2h2-classical-velocity}). The pulse’s effect is only evident in Fig.~\ref{fig:c2h2-classical-pulse-velocity} between 10 and 30 fs, where each ion’s velocity exhibits slight oscillations during the pulse duration (Fig.~\ref{fig:pulse-electron-c2h2}). This oscillation is minimal, with peak fluctuations not exceeding an amplitude of 0.05 \AA/fs, resulting in negligible position changes, hence the apparent smoothness in Fig.~\ref{fig:c2h2-classical-pulse-positions}. Additionally, the electric field’s symmetric oscillations produce zero net impulse, explaining why it has zero effect on the positions and velocities beyond the pulse duration. This reasoning also applies to the semi-classical simulations, as shown in the nearly identical ion positions with and without the pulse (Fig.~\ref{fig:c2h2-semi-classical-pulse-positions} and Fig.~\ref{fig:c2h2-semi-classical-positions}) and the velocities (Fig.~\ref{fig:c2h2-semi-classical-pulse-velocity} and Fig.~\ref{fig:c2h2-semi-classical-velocity}).

The quantum ion positions and velocities are shown in Fig.~\ref{fig:c2h2-quantum-positions} and Fig.~\ref{fig:c2h2-quantum-velocity}. As seen in Fig.~\ref{fig:c2h2-quantum-positions}, the ions do not begin to repel each other until ionization starts, as expected. During ionization, the atoms’ velocities oscillate due to the alternating electric field of the laser, which periodically pushes and pulls the ions as the field reverses direction. This results in velocity oscillations alongside a general increase in speed due to the Coulomb repulsion of ions, as shown in Fig.~\ref{fig:c2h2-quantum-velocity}.

Several key observations emerge when comparing the classical kinematic graphs (Fig.~\ref{fig:c2h2-classical-positions} and Fig.~\ref{fig:c2h2-classical-velocity}), semi-classical kinematic graphs (Fig.~\ref{fig:c2h2-semi-classical-positions} and Fig.~\ref{fig:c2h2-semi-classical-velocity}), and quantum kinematic graphs (Fig.~\ref{fig:c2h2-quantum-positions} and Fig.~\ref{fig:c2h2-quantum-velocity}). Two notable conclusions, along with their explanations, are as follows:

First, the classical model predicts both a greater final displacement from the origin and a higher final velocity for all atoms at the end of the simulation window (80~fs) compared to the quantum model. This discrepancy arises because, in the classical model, the atoms continuously repel each other based on their final charge states throughout the entire 80~fs simulation. In contrast, the quantum model establishes the final charge states only after ionization is complete, around \( t^* \). As a result, Coulombic repulsion at the final charge levels in the quantum model occurs for only approximately 55~fs, starting after \( t^* \). Once the laser begins to ionize the molecule (at around 12~fs in Fig.~\ref{fig:pulse-electron-c2h2}), the ions initially carry very little charge, resulting in a weaker repulsion between them. This repulsion gradually increases as the ionization progresses, reaching its maximum strength at \( t^* \), when the ions achieve their final and maximum charge states for the strongest Coulombic repulsion.

Second, the quantum and semi-classical models predict nearly identical position and velocity curves after ionization. This suggests that, once ionization is complete, the atomic repulsion follows a behavior closely resembling the classical model. This implies that quantum effects play a negligible role at larger distances when the atoms are further separated as expected.

\begin{figure*}
\centering
\includegraphics[width=\textwidth]{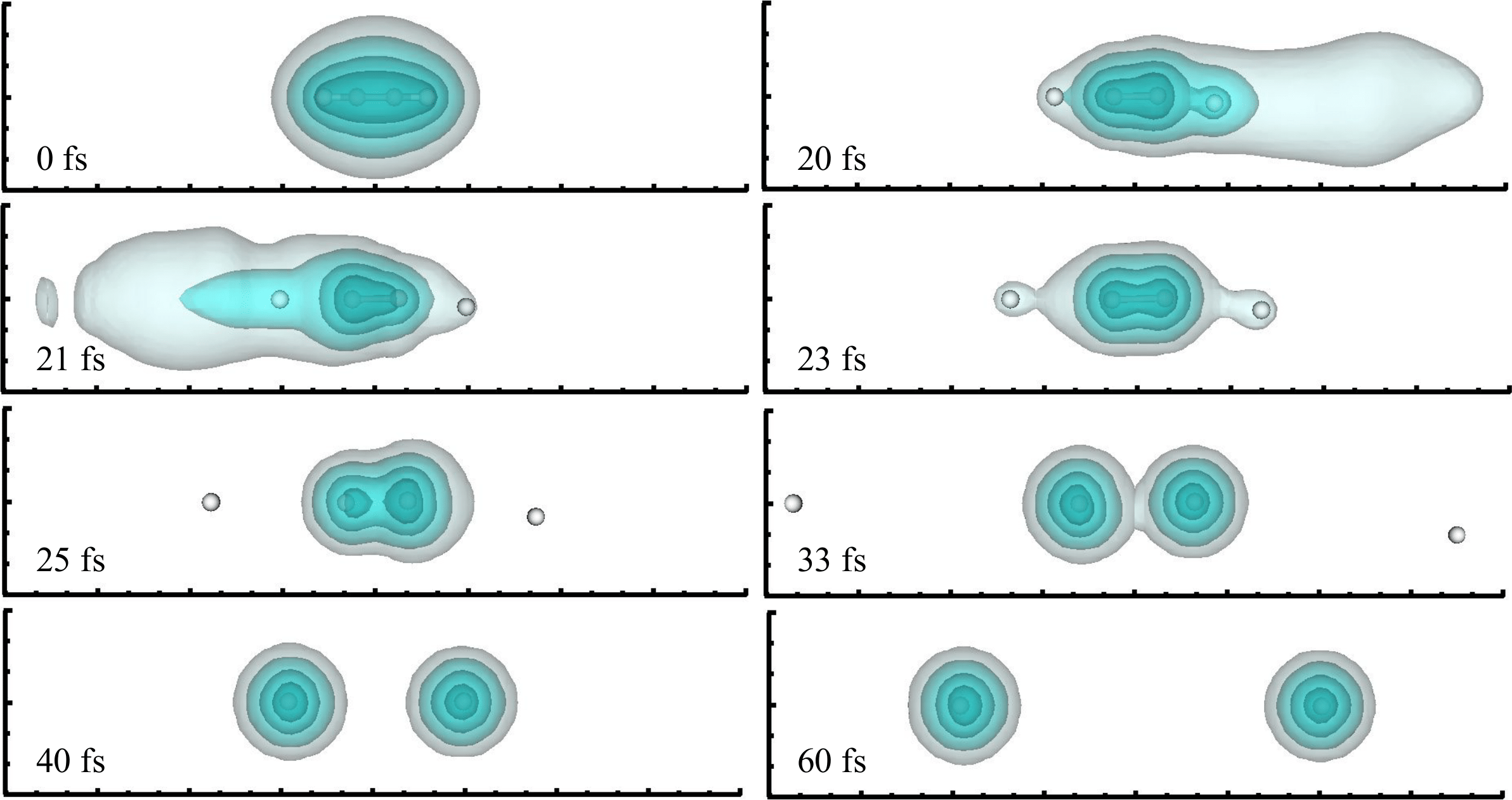}
\caption{Coulomb explosion snapshots of one TDDFT quantum simulation. Axes are marked with tick marks at intervals of 1~\AA. The 0.5, 0.1, 0.01, and 0.001 density isosurfaces are shown.}
\label{fig:quantum-snapshot}
\end{figure*}

\begin{figure*}
\centering
\includegraphics[width=\textwidth]{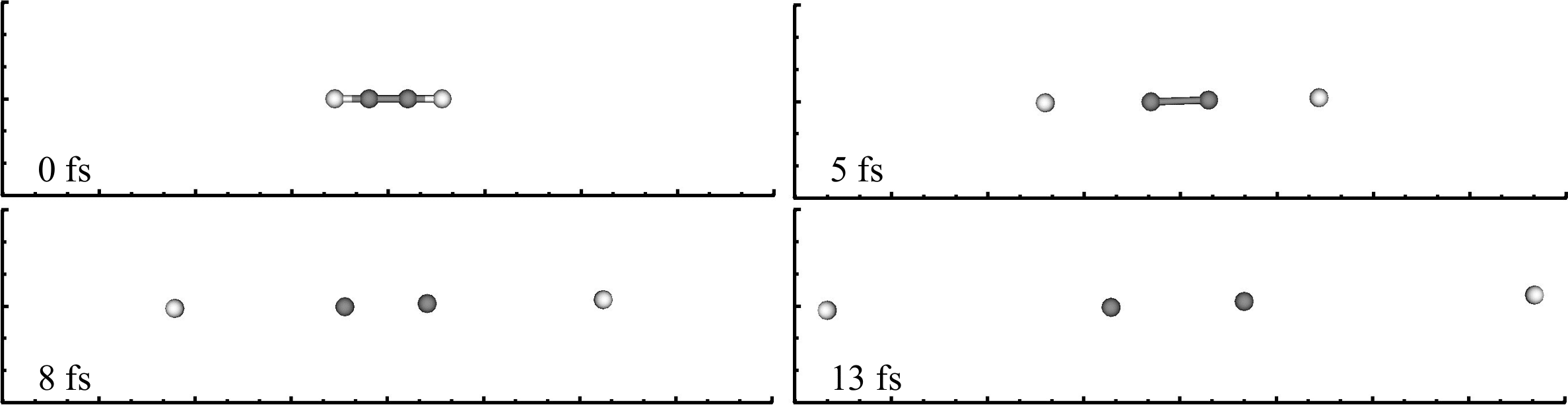}
\caption{Coulomb explosion snapshots of one classical simulation. Axes are marked with tick marks at intervals of 1~\AA.}
\label{fig:classical-snapshot}
\end{figure*}

Snapshots from a quantum TDDFT simulation are shown in Fig.~\ref{fig:quantum-snapshot}, illustrating the evolution of electron and ion dynamics during and after ionization. In the figure, the hydrogen atoms are ejected early in the ionization process, reaching approximately 2~\AA\ from their respective carbon atoms by 21~fs and approximately 3~\AA\ by 23~fs. At 25~fs (\( t^* \) in the semi-classical simulation, marking the completion of ionization), the carbon atoms begin to repel at full force, and the hydrogen atoms are displaced by nearly 4~\AA. These observations highlight two key dynamics in the system: (1) hydrogen atoms lose their electrons first and begin repelling during ionization, and (2) by the end of ionization, the hydrogen atoms are significantly separated from the system (about 4~\AA\ from the nearest carbon), and the carbon atoms are also separating (about 2~\AA\ apart). As these interatomic and electron distances increase further, quantum effects are expected to become increasingly negligible.

Fig.~\ref{fig:classical-snapshot} complements Fig.~\ref{fig:quantum-snapshot} by showing the ion positions from a classical simulation. In this case, the atoms undergo rapid repulsion from the start of the simulation, with the lighter hydrogen atoms being ejected first. By 13~fs, the atoms are already significantly separated, with the hydrogen atoms approximately 9~\AA\ from the nearest carbon and the carbon atoms about 4~\AA\ apart. These distances are comparable to those observed in Fig.~\ref{fig:quantum-snapshot} at 33~fs.

\subsection{Acetylene Distributions}

\begin{figure*}[ht!]
    \centering
    \includegraphics[width=\textwidth]{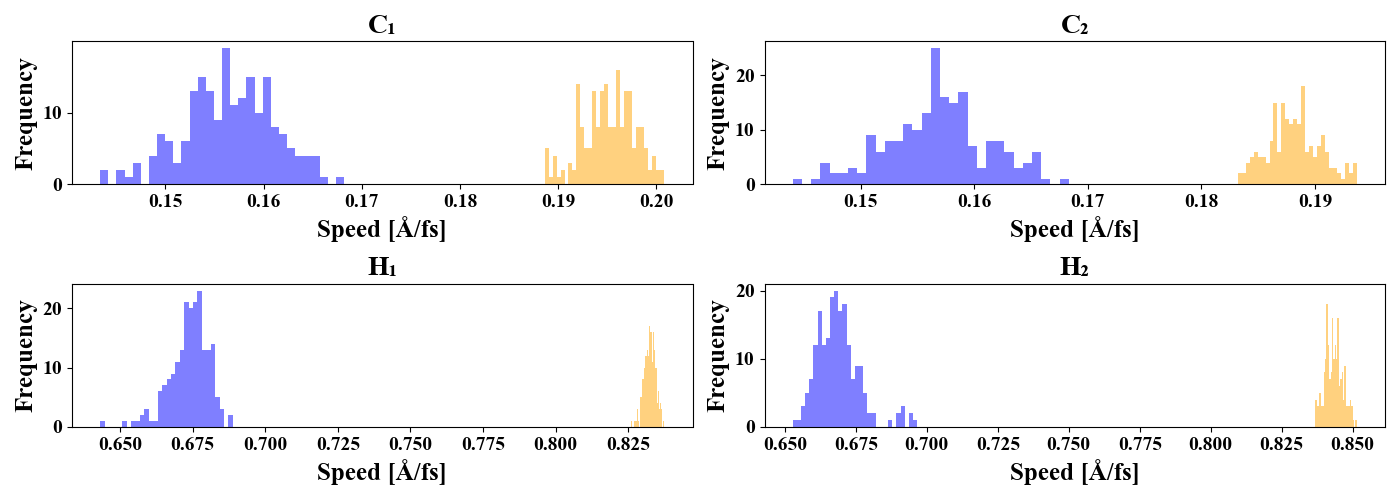}
    \caption{Final speed and frequency histogram for each of the ions in C\textsubscript{2}H\textsubscript{2} Coulomb explosion. Purple is the final velocities for all of the atoms in the quantum simulations. Orange is the final velocities for all of the atoms in the classical simulations.}
    \label{fig:c2h2-speed-hist}
\end{figure*}

\begin{figure*}[ht!]
    \centering
    \begin{subfigure}[b]{0.49\textwidth}
        \centering
        \includegraphics[width=\textwidth]{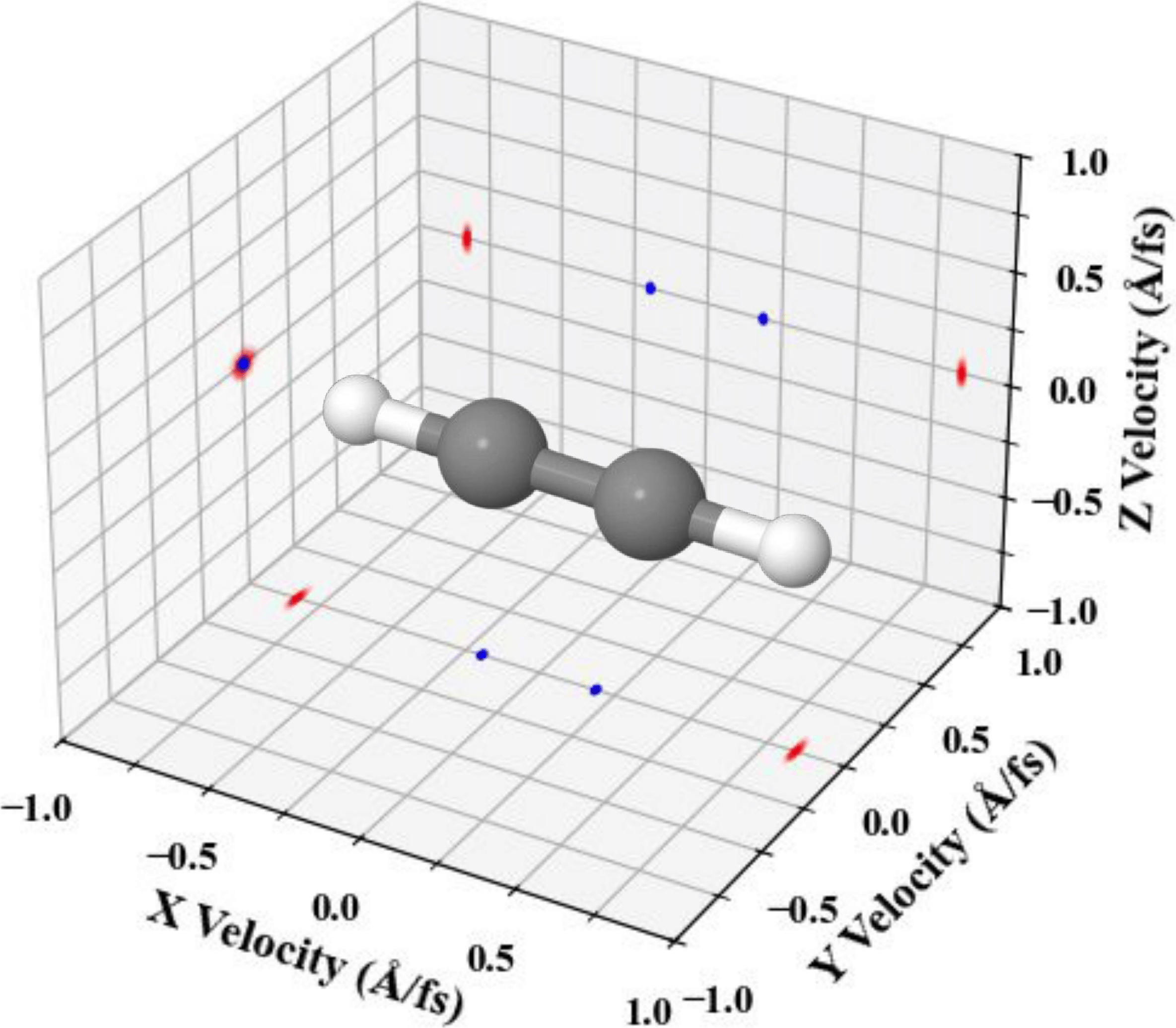}
        \caption{Classical method Newton plot}
        \label{fig:c2h2-classical-projection}
    \end{subfigure}
    \hfill
    \begin{subfigure}[b]{0.49\textwidth}
        \centering
        \includegraphics[width=\textwidth]{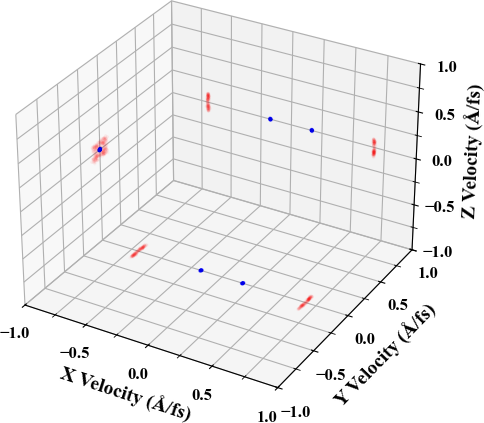}
        \caption{Semi-classical method Newton plot}
        \label{fig:c2h2-semi-classical-projection}
    \end{subfigure}

    \vspace{0.5cm}
    
    \begin{subfigure}[b]{0.49\textwidth}
        \centering
        \includegraphics[width=\textwidth]{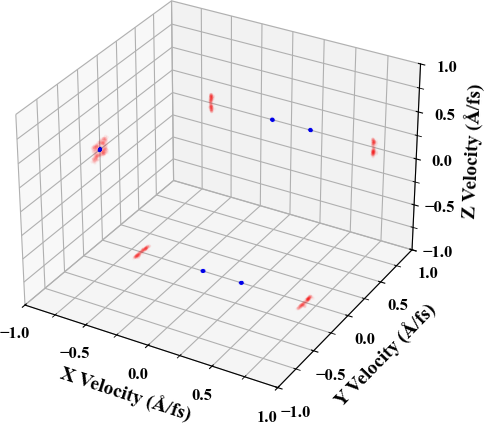}
        \caption{Quantum method Newton plot}
        \label{fig:c2h2-quantum-projection}
    \end{subfigure}
    
    \caption{3D final velocity projection plots of each carbon (blue) and hydrogen (red) ion resulting from Coulomb explosion of C\textsubscript{2}H\textsubscript{2}.}
    \label{fig:c2h2-newton-plots-projection}
\end{figure*}

\begin{figure*}[ht!]
    \centering
    \begin{subfigure}[b]{0.49\textwidth}
        \centering
        \includegraphics[width=\textwidth]{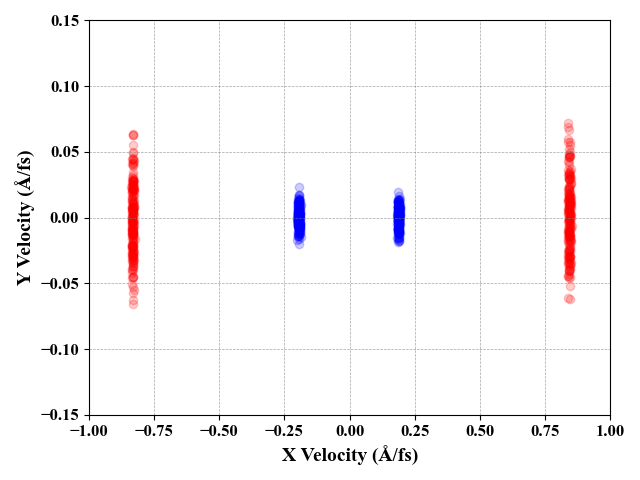}
        \caption{Classical method}
        \label{fig:c2h2-classical-xy-projection}
    \end{subfigure}
    \hfill
    \begin{subfigure}[b]{0.49\textwidth}
        \centering
        \includegraphics[width=\textwidth]{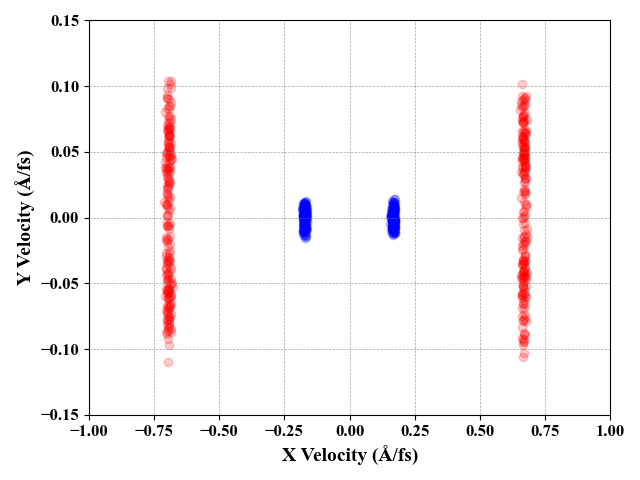}
        \caption{Semi-classical method}
        \label{fig:c2h2-semi-classical-xy-projection}
    \end{subfigure}

    \vspace{0.5cm}
    
    \begin{subfigure}[b]{0.49\textwidth}
        \centering
        \includegraphics[width=\textwidth]{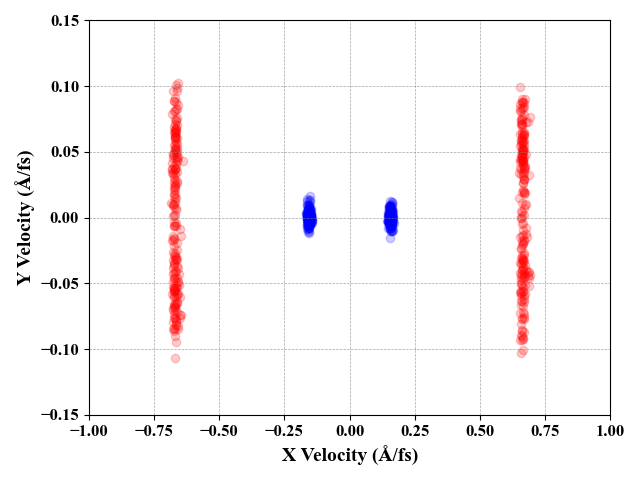}
        \caption{Quantum method}
        \label{fig:c2h2-quantum-xy-projection}
    \end{subfigure}
    
    \caption{2D xy-projection Newton plots of the final velocity of each carbon (blue) and hydrogen (red) atom resulting from Coulomb explosion of C\textsubscript{2}H\textsubscript{2}.}
    \label{fig:c2h2-xy-projections}
\end{figure*}

\begin{figure*}[ht!]
    \centering
    \begin{subfigure}[b]{0.49\textwidth}
        \centering
        \includegraphics[width=\textwidth]{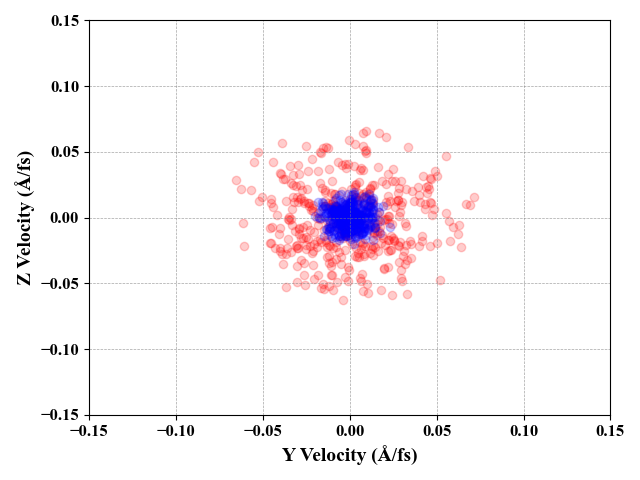}
        \caption{Classical method}
        \label{fig:c2h2-classical-yz-projection}
    \end{subfigure}
    \hfill
    \begin{subfigure}[b]{0.49\textwidth}
        \centering
        \includegraphics[width=\textwidth]{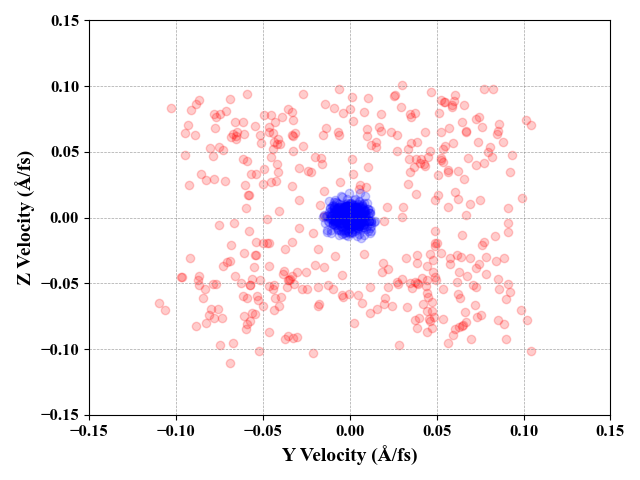}
        \caption{Semi-classical method}
        \label{fig:c2h2-semi-classical-yz-projection}
    \end{subfigure}

    \vspace{0.5cm}
    
    \begin{subfigure}[b]{0.49\textwidth}
        \centering
        \includegraphics[width=\textwidth]{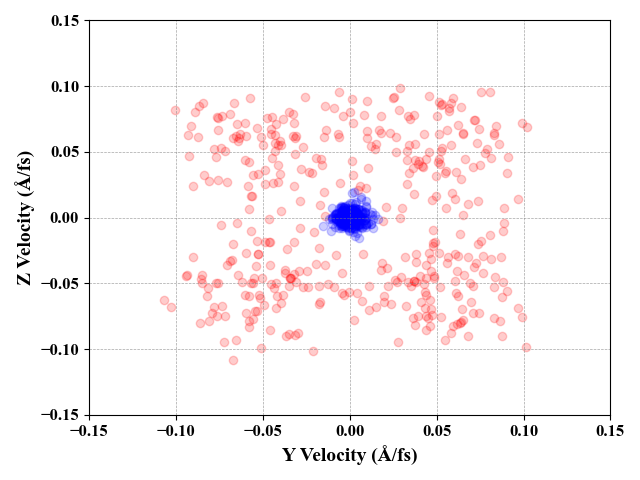}
        \caption{Quantum method}
        \label{fig:c2h2-quantum-yz-projection}
    \end{subfigure}
    
    \caption{2D yz-projection plots of the final velocity of each carbon (blue) and hydrogen (red) atom resulting from Coulomb explosion of C\textsubscript{2}H\textsubscript{2}.}
    \label{fig:c2h2-yz-projections}
\end{figure*}

\begin{figure*}[ht!]
    \centering
    \begin{subfigure}[b]{0.49\textwidth}
        \centering
        \includegraphics[width=\textwidth]{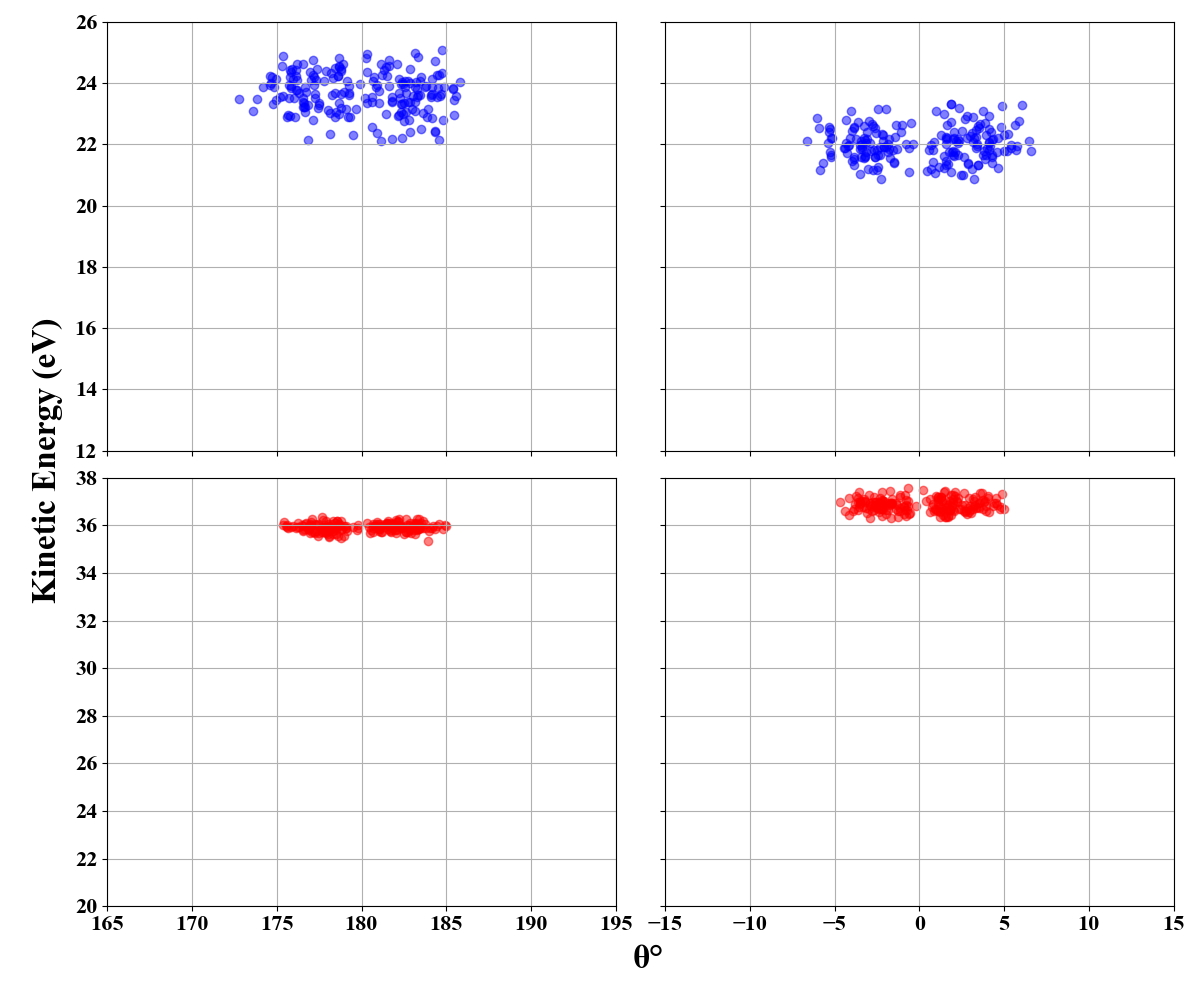}
        \caption{Classical method}
        \label{fig:c2h2-classical-angular-dist}
    \end{subfigure}
    \hfill
    \begin{subfigure}[b]{0.49\textwidth}
        \centering
        \includegraphics[width=\textwidth]{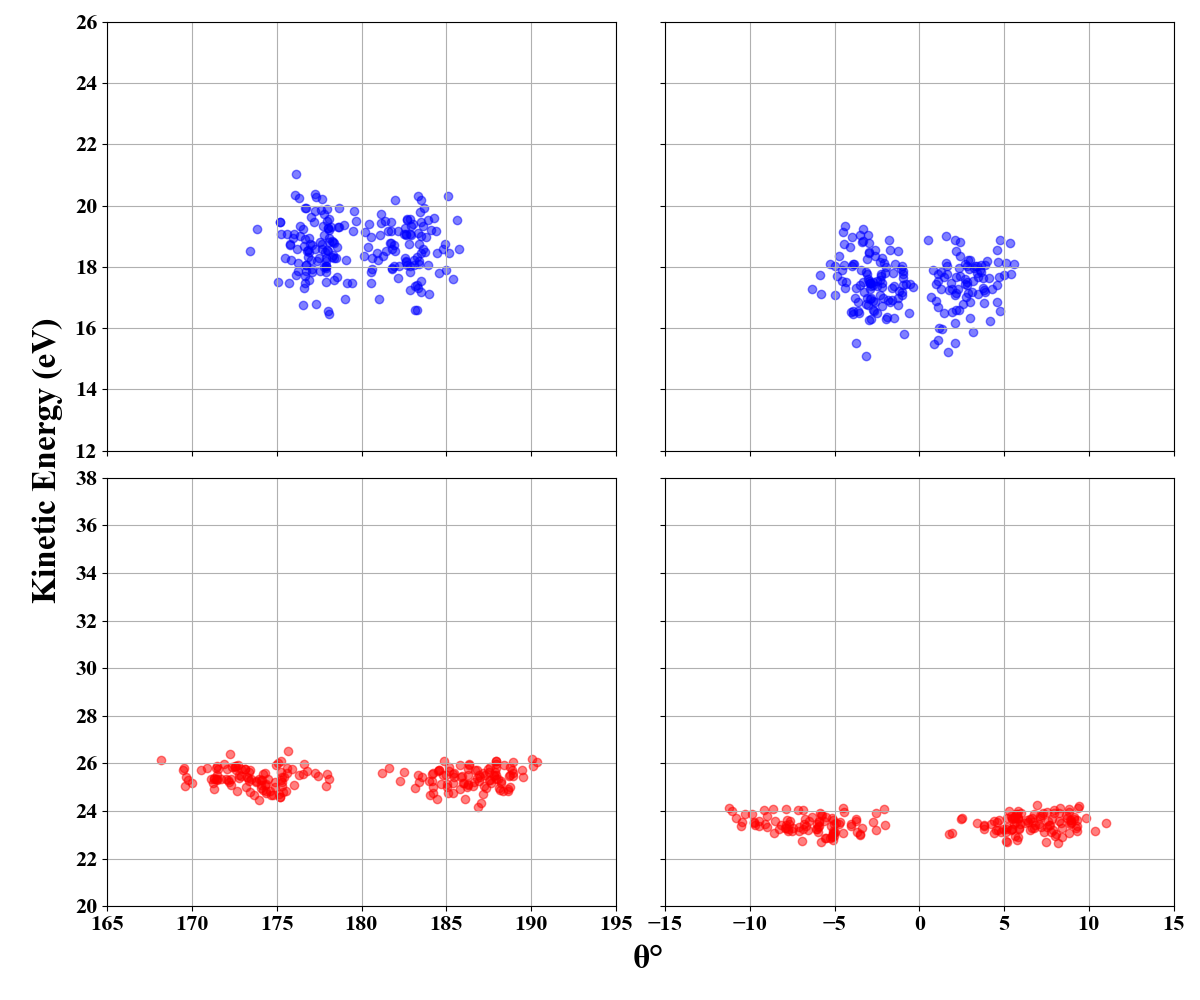}
        \caption{Semi-classical method}
        \label{fig:c2h2-semi-classical-angular-dist}
    \end{subfigure}

    \vspace{0.5cm}
    \begin{subfigure}[b]{0.49\textwidth}
        \centering
        \includegraphics[width=\textwidth]{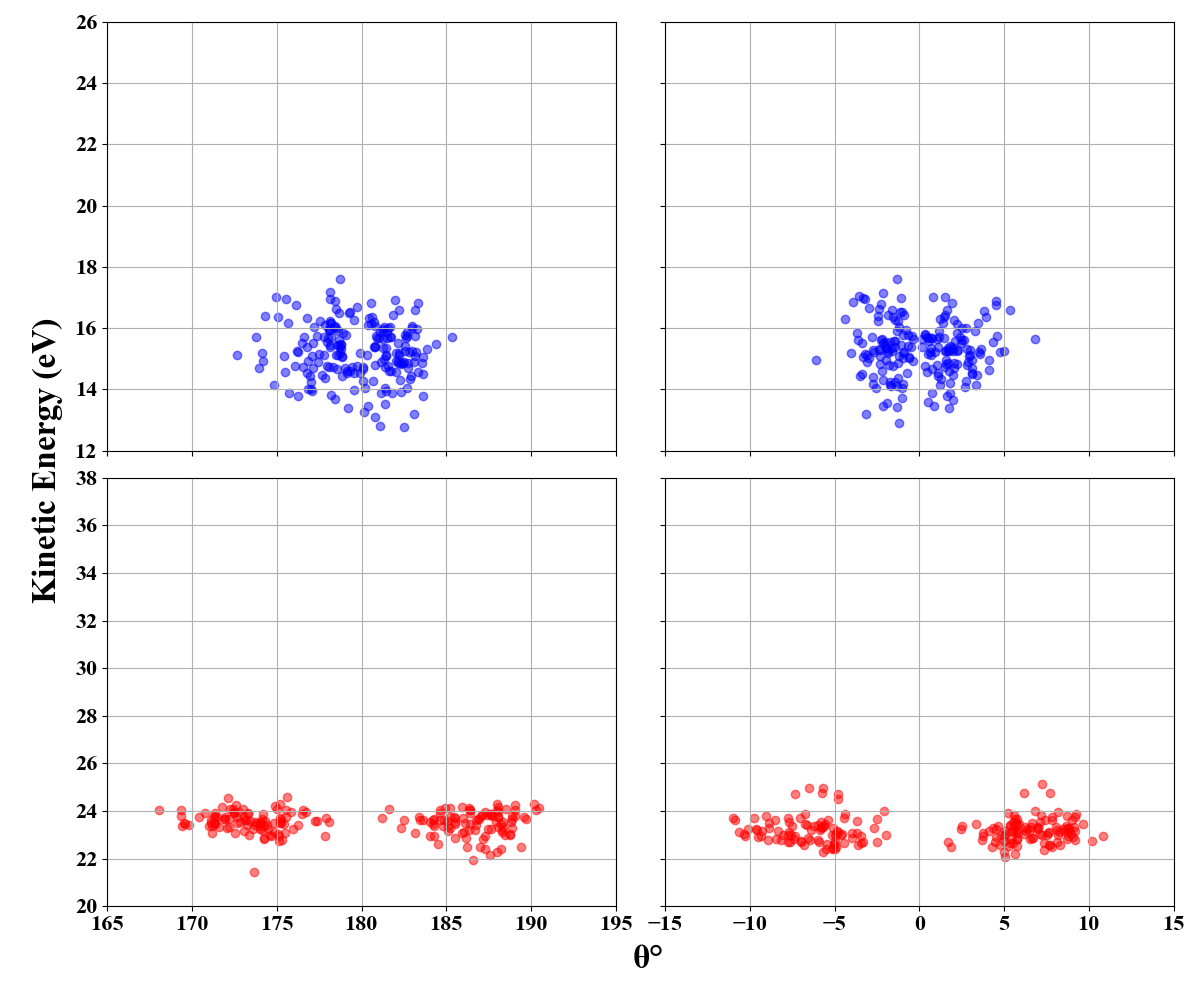}
        \caption{Quantum method}
        \label{fig:c2h2-quantum-angular-dist}
    \end{subfigure}
    \hfill
    \begin{subfigure}[b]{0.49\textwidth}
        \centering
        \includegraphics[width=\textwidth]{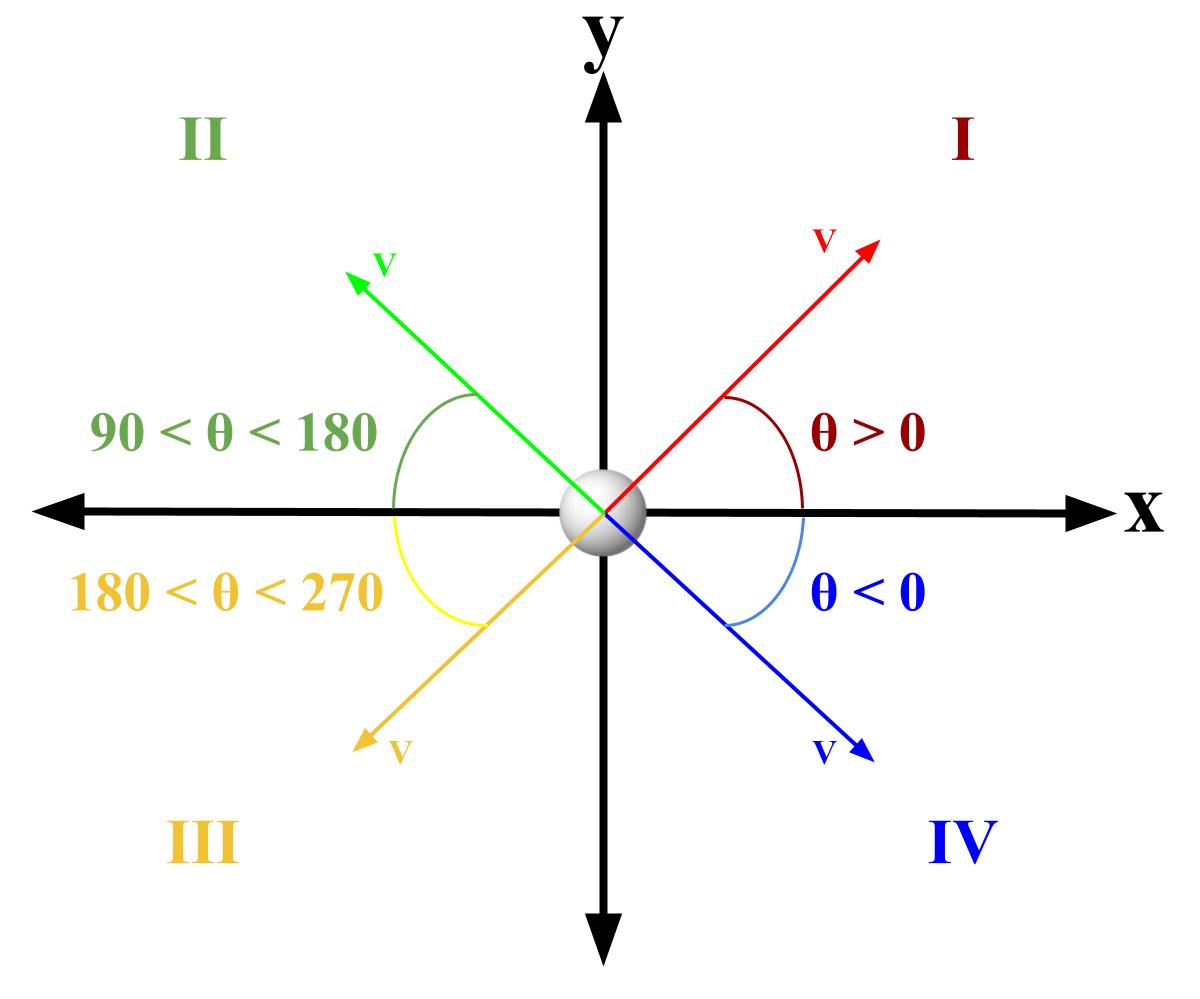}
        \caption{Diagram of the values for $\theta$}
        \label{fig:c2h2-angular-theta-diagram}
    \end{subfigure}
    
    \caption{Angular distribution and final kinetic energy of each carbon (blue) and hydrogen (red) atom resulting from Coulomb explosion of C\textsubscript{2}H\textsubscript{2}. $\theta$ is the angle of the velocity vector relative to the x-axis.}
    \label{fig:c2h2-angular-distribution}
\end{figure*}

To ensure accurate comparisons of final velocities across all models, all figures representing an ion's final speed or kinetic energy (Fig.~\ref{fig:c2h2-speed-hist}–\ref{fig:isoxazole-newton-plots-projection})
 use data collected from the same spatial point during the simulation. Each ion’s \enquote{final velocity} is recorded when it reaches a tolerance distance of 1.5 \AA \, from the CAP boundary set in the quantum simulation. Although ions may reach this region at different times, this approach ensures that each ion has traveled the same spatial distance before reaching the measurement point, enabling consistent comparisons across methods. Collecting final kinetic information at this point also guarantees that, in the quantum case, data are recorded prior to any CAP-induced electron loss, ensuring that all measurements reflect pure quantum effects.

Fig.~\ref{fig:c2h2-speed-hist} illustrates the differences in the final velocities of each atom in C\textsubscript{2}H\textsubscript{2} between the classical and quantum models. As shown in the figure, by the time the atoms reach the complex absorbing potential, the classical model predicts higher velocities for all atoms. Since both models involve the same initial distance to traverse from its near equilibrium geometry to the CAP, this velocity difference arises from the quantum effects, presence of electrons densities in the system, and how ionization is modeled. The electron cloud surrounding the carbon atoms generates additional attractive forces that slow the hydrogen atoms' trajectories. Similarly, the carbon atoms experience a reduction in velocity due to the attractive forces between their positive nuclei and the surrounding electrons. Additionally, because the atoms begin to repel each other before reaching their maximum charge states in the quantum model, their final velocities are reduced.

Coulomb explosion imaging experiments \cite{PhysRevLett.132.123201,PhysRevResearch.4.013029}
can measure ion momenta and the ion distribution in the three dimensional fragment-ion momentum space
(Newton plots). These distribution patterns give direct information about the molecular structures.
Fig.~\ref{fig:c2h2-newton-plots-projection} shows Newton plots for each method of C\textsubscript{2}H\textsubscript{2} Coulomb explosion, illustrating the 3-dimensional projections of the final ion velocities on each coordinate plane. Each ion in each simulation is represented by a transparent circle, with hydrogen shown in red and carbon in blue. By plotting each ion in every simulation, all three simulation methods generate a distinctive distribution that reflects the molecular structure of C\textsubscript{2}H\textsubscript{2}, as would be expected experimentally where quantum effects are at play~\cite{PhysRevLett.132.123201}. However, the precision with which each method represents the structure varies (the actual equilibrium geometry of the C\textsubscript{2}H\textsubscript{2} molecule is overlaid for comparison in Fig.~\ref{fig:c2h2-classical-projection}). The classical method predicts a precise and narrower spread with a more idealized set of final ion velocities compared to the quantum method, while the semi-classical method falls between these two extremes. 

The spread of velocity data points is further illustrated in the 2D projections onto individual planes (see Fig.~\ref{fig:c2h2-xy-projections} and Fig.~\ref{fig:c2h2-yz-projections}). In the classical model (Fig.~\ref{fig:c2h2-classical-xy-projection}), the y-velocities of hydrogen ions are confined to a narrower range, spanning from -0.07~\AA/fs to 0.07~\AA/fs. In contrast, the quantum model (Fig.~\ref{fig:c2h2-quantum-xy-projection}) exhibits a broader distribution, with y-velocities ranging from -0.11~\AA/fs to 0.10~\AA/fs. This observation suggests that, although the classical model predicts higher maximum hydrogen velocities along the x-axis (approximately -0.83~\AA/fs to 0.83~\AA/fs) compared to the quantum model (which predicts x-velocities ranging from -0.68~\AA/fs to 0.68~\AA/fs), the classical model results in a significantly narrower overall velocity distribution for the ejected hydrogen ions. This highlights the enhanced variability in ion dynamics captured by the quantum model.

This trend\textemdash smaller angular distribution for hydrogen in the classical model\textemdash remains consistent when comparing the classical and quantum 2D velocity projections onto the yz-plane (Fig.~\ref{fig:c2h2-yz-projections}). The classical model significantly underestimates the spread of hydrogen atoms compared to the quantum model. Notably, in both the xy- and yz-plane projections, the semi-classical method (Figs.~\ref{fig:c2h2-semi-classical-xy-projection} and \ref{fig:c2h2-semi-classical-yz-projection}) shows strong agreement with the quantum method (Figs.~\ref{fig:c2h2-quantum-xy-projection} and \ref{fig:c2h2-quantum-yz-projection}). This strong alignment between the semi-classical and quantum methods further suggests that the discrepancies between the classical and quantum models arise when the ions are in close proximity, where quantum effects have a significant influence, and during ionization.

The narrower spread observed in the classical Newton plots compared to the quantum plots (Fig.~\ref{fig:c2h2-newton-plots-projection}–\ref{fig:c2h2-yz-projections}) suggests that quantum effects and the presence of electrons not only reduce the overall velocity of the ions but also increase the diversity of simulation outcomes. This behavior arises from the attractive forces exerted by the electrons on the ions, as well as the additional repulsive interactions between the ions and the electrons of neighboring atoms. These interactions can significantly alter the trajectories and velocities of the ions in various ways, resulting in a broader range of possible outcomes.

The angular distribution of each ion across all C\textsubscript{2}H\textsubscript{2} Coulomb explosion simulations is displayed in Fig.~\ref{fig:c2h2-angular-distribution}. Here, the y-axis shows the ions’ final kinetic energy, while the x-axis represents the angle $\theta$ of each ion's velocity vector relative to the x-axis. The domain for $\theta$ values is defined in Fig.~\ref{fig:c2h2-angular-theta-diagram}, where $\theta$'s domain corresponds to the projection of the velocity on the xy-plane. Specifically, $\theta$ is positive if the x- and y-components of the velocity vector, $v_x$ and $v_y$, are both positive, placing the vector in Quadrant I of the xy-plane. In Quadrant II, where $v_x$ is negative and $v_y$ positive, $90^\circ < \theta < 180^\circ$. In Quadrant III, where $v_x$ and $v_y$ are negative, $180^\circ < \theta < 270^\circ$. Finally, in Quadrant IV, when when $v_x$ is positive and $v_y$ is negative, $\theta$ is negative, as shown in Fig.~\ref{fig:c2h2-angular-theta-diagram}.

This method of assigning $\theta$ values ensures a continuous and linear transition of angles along the x-axis as they cross the y-axis. All ions have trajectories that span both Quadrants I and II, while others extend into Quadrants III and IV (see Fig.~\ref{fig:c2h2-classical-angular-dist}, Fig.~\ref{fig:c2h2-semi-classical-angular-dist}, and Fig.~\ref{fig:c2h2-quantum-angular-dist}).

By examining the angular distributions from each simulation method shown in Fig.~\ref{fig:c2h2-angular-distribution}, several insights emerge. Notably, the angular distributions (x-axis) of carbon ions in both the classical and quantum methods are surprisingly similar. Both methods predict nearly identical angular ranges for the carbon ions: from $173^\circ$ to $186^\circ$ for the left-side carbon and from $-7^\circ$ to $7^\circ$ for the right-side carbon, with comparable data spreads within these ranges. However, significant differences appear in the kinetic energy distributions (y-axis). In the quantum method, the kinetic energy of the carbon ions ranges from approximately 12.8~eV to 17.8~eV, spanning a total of 5~eV. In contrast, the classical method predicts a narrower kinetic energy spread, with the left carbon atom varying from about 22.1~eV to 25.1~eV and the right carbon atom from 20.9~eV to 23.3~eV, yielding a maximum spread of 3~eV. The broader kinetic energy range in the quantum method suggests a greater variety of possible outcomes, attributed to quantum effects, the detailed inclusion of electron interactions, and the ionization process. This larger spread of kinetic energies is also evident in Fig.~\ref{fig:c2h2-speed-hist}, where the quantum method data exhibits a more diffuse spread compared to the classical method data.

The angular distributions and kinetic energies of the hydrogen atoms exhibit even greater differences between the classical and quantum models. In the classical method, the left hydrogen's kinetic energy ranges from 35.4~eV to 36.4~eV, and the right hydrogen's from 36.3~eV to 37.6~eV, resulting in a total spread of only 1.4~eV. In contrast, the quantum method predicts a left hydrogen energy range of 21.4~eV to 24.6~eV and a right hydrogen range of 22.1~eV to 25.1~eV, giving a much larger spread of 3.2~eV. These results not only highlight the significantly higher final kinetic energy in the classical model (approximately 36~eV) compared to the quantum model (approximately 23~eV) but also reveal the more constrained range of possible outcomes in the classical case. 
This trend\textemdash of the classical model providing less diverse range of atom kinetic energies\textemdash extends to angular distributions as well. In the classical method, the hydrogen angular distributions are confined to a range of $10^\circ$, whereas the quantum method predicts a broader range of $22^\circ$, spanning $168^\circ$ to $190^\circ$ for the left hydrogen and $-11^\circ$ to $11^\circ$ for the right hydrogen. These findings emphasize the limited variability in velocities and angular distributions in the classical simulation, underscoring its idealized nature and the significant role quantum effects play in introducing variability and realism to the results.

The semi-classical method's angular distributions (Fig.~\ref{fig:c2h2-semi-classical-angular-dist}) fall between those of the quantum method (Fig.~\ref{fig:c2h2-quantum-angular-dist}) and the classical method (Fig.~\ref{fig:c2h2-classical-angular-dist}). Generally, the semi-classical approach closely aligns with the quantum method in terms of range and diversity of outcomes, particularly because strong quantum effects dominate before the atoms become significantly separated. However, the representation of atoms as point charges after $t^*$ increases their final kinetic energies compared to the quantum method. These observations lead to the conclusion that quantum effects in Coulomb explosion not only result in broader angular distributions and greater diversity of outcomes but also reduce the final velocities of the ejected atoms.

\subsection{Butane (C\textsubscript{4}H\textsubscript{10}) Distributions}

\begin{figure*}[ht!]
    \centering
    \begin{subfigure}[b]{0.49\textwidth}
        \centering
        \includegraphics[width=\textwidth]{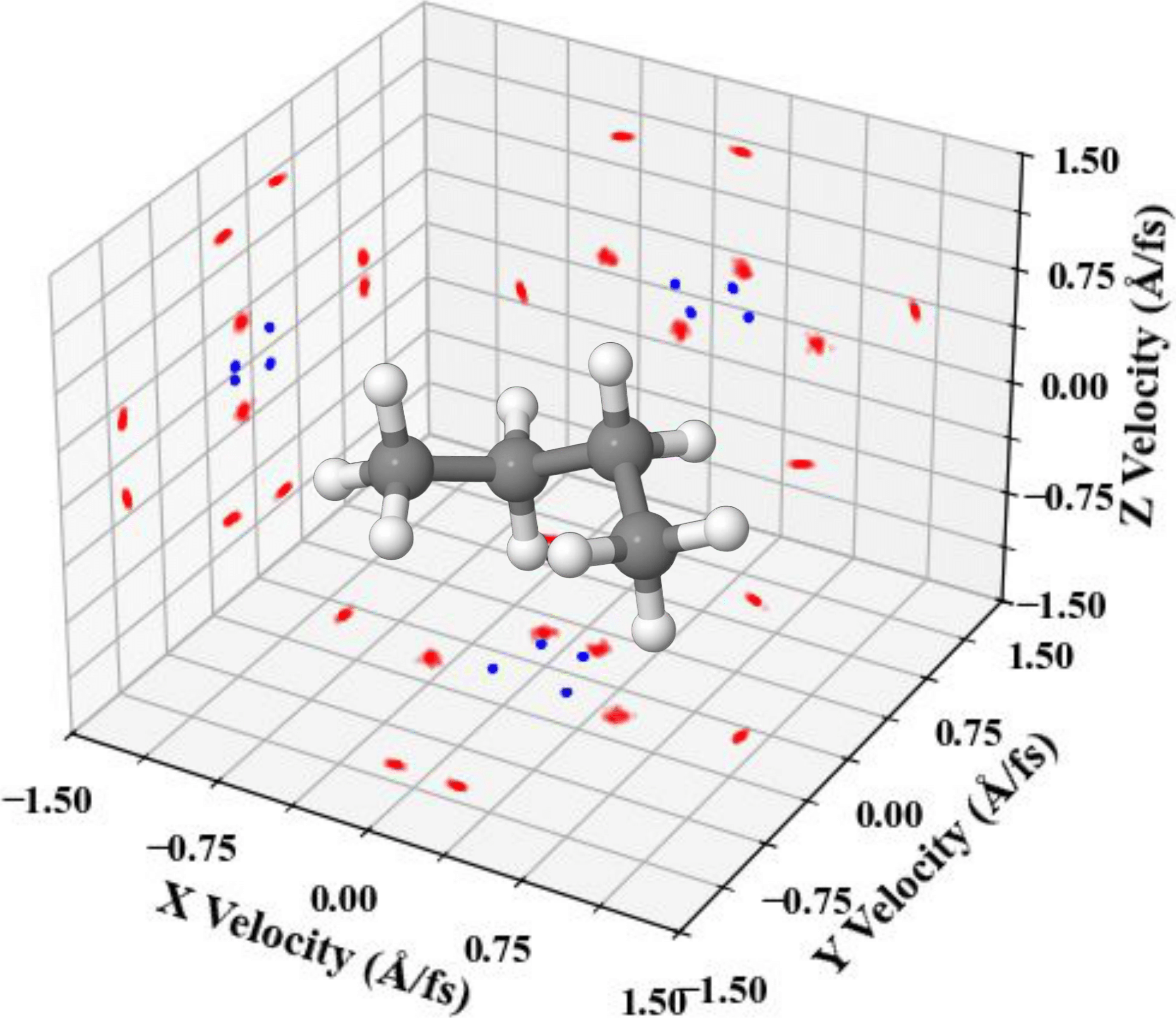}
        \caption{Classical method (14 V/\AA\ maximum E-field)}
        \label{fig:c4h10-14-classical-projection}
    \end{subfigure}
    \hfill
    \begin{subfigure}[b]{0.49\textwidth}
        \centering
        \includegraphics[width=\textwidth]{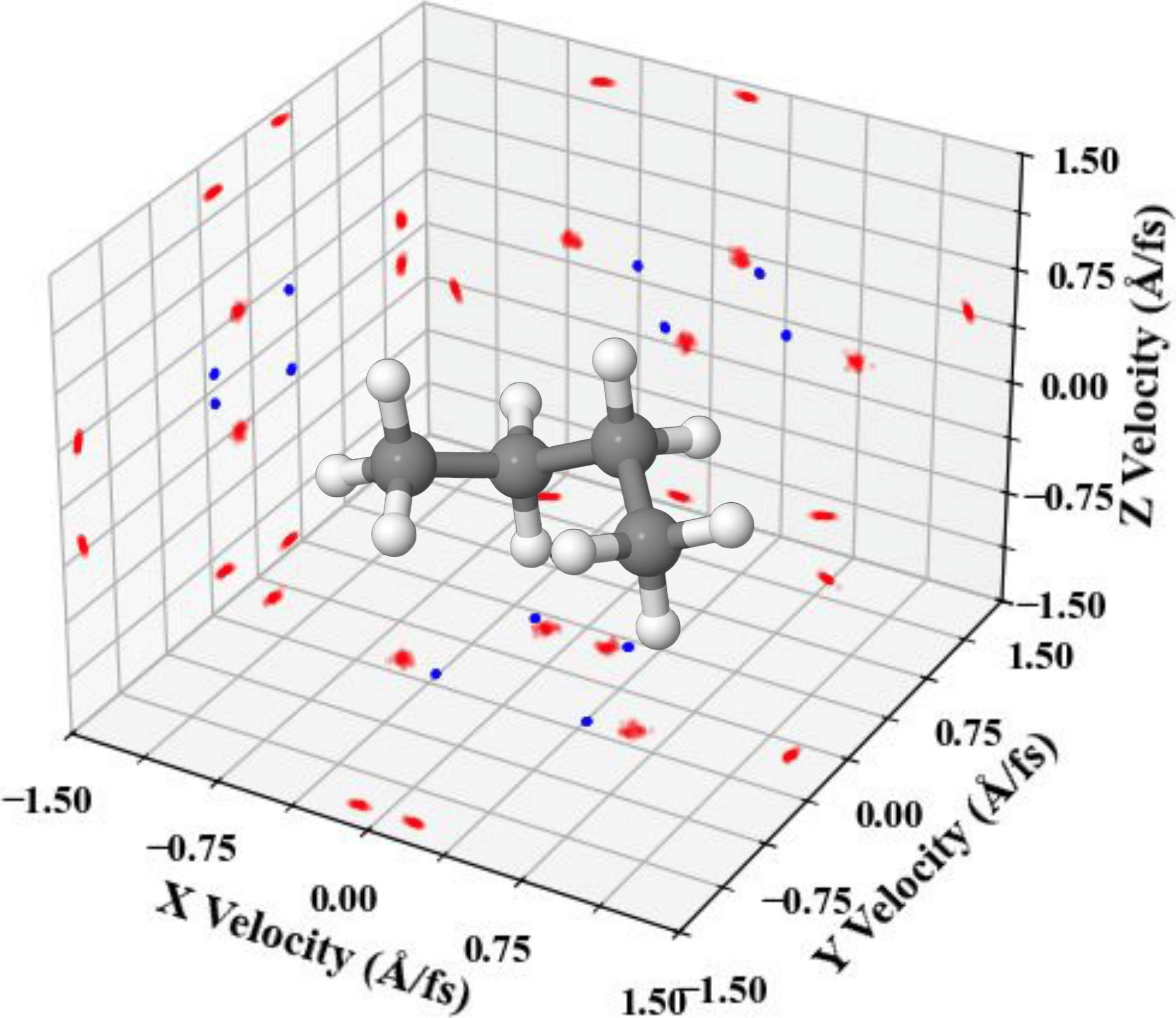}
        \caption{Classical method (28 V/\AA\ maximum E-field)}
        \label{fig:c4h10-28-classical-projection}
    \end{subfigure}
    
    \vspace{0.5cm}
    
    \begin{subfigure}[b]{0.49\textwidth}
        \centering
        \includegraphics[width=\textwidth]{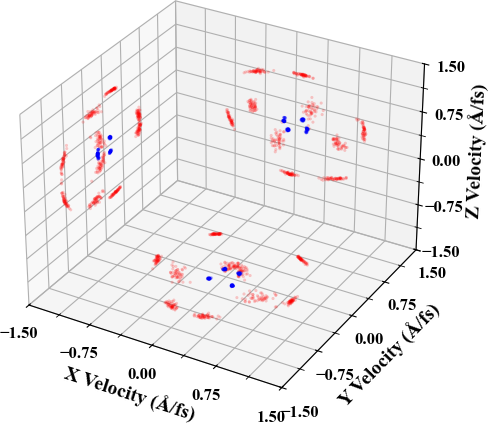}
        \caption{Quantum method (14 V/\AA\ maximum E-field)}
        \label{fig:c4h10-14-quantum-projection}
    \end{subfigure}
    \hfill
    \begin{subfigure}[b]{0.49\textwidth}
        \centering
        \includegraphics[width=\textwidth]{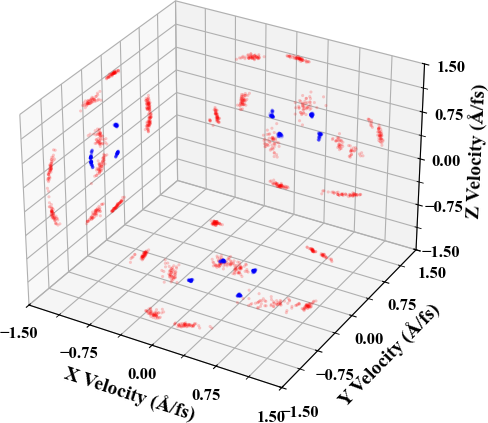}
        \caption{Quantum method (28 V/\AA\ maximum E-field)}
        \label{fig:c4h10-28-quantum-projection}
    \end{subfigure}

    \caption{3D final velocity projection Newton plots of each carbon (blue) and hydrogen (red) atom resulting from Coulomb explosion of C\textsubscript{4}H\textsubscript{10}.}
    \label{fig:c4h10-newton-plots-projections}
\end{figure*}

Applying this comparative approach across different molecules reveals consistent patterns: the classical method consistently produces narrower angular distributions, less diverse trajectory outcomes, and higher velocities for all atoms. These trends are particularly apparent in the Newton plot projections of the Coulomb explosion of butane in its gauche conformation, as shown in Fig.~\ref{fig:c4h10-newton-plots-projections}. For these figures, 50 simulations of Coulomb explosion were conducted on C\textsubscript{4}H\textsubscript{10} using the pulse depicted in Fig.~\ref{fig:pulse-electron-c2h2}, and an additional 50 simulations were performed with the maximum electric field amplitude doubled from 14 V/\AA\ to 28 V/\AA. 

Fig.~\ref{fig:c4h10-newton-plots-projections} highlights the distinct differences between classical and quantum methods of Coulomb explosion, as well as the impact of using a stronger electric field. In the classical simulations for the 14 V/\AA\ field, each atom was assigned an average charge based on the results of the quantum simulations: carbon atoms were set to charges of $1.9^{+}$, $1.8^{+}$, $1.8^{+}$, and $2.0^{+}$, while each hydrogen atom was assigned a charge of $1^{+}$. For the simulations with the 28 V/\AA\ field, the carbon charges were set to $3.9^{+}$ and the hydrogen to $1^{+}$, reflecting the increased ionization. For the quantum simulations of butane, the CAP was set at -10 and 10 \AA\ along the x-axis, and -9 and 9 \AA\ along the y- and z-axes.

The classical Newton plots for butane subjected to electric fields of 14~V/\AA\ (Fig.~\ref{fig:c4h10-14-classical-projection}) and 28~V/\AA\ (Fig.~\ref{fig:c4h10-28-classical-projection}) exhibit a similar overall structure, with the primary distinction being the greater displacement of all ions from the origin in the 28~V/\AA\ case. This difference is expected, as the only variable altered between the two simulations is the charge assigned to the carbon atoms. The increased charge in the 28~V/\AA\ case leads to stronger Coulombic repulsion, resulting in higher velocities, as reflected in Fig.~\ref{fig:c4h10-newton-plots-projections}.

The quantum Newton plots for butane under a 14~V/\AA\ electric field (Fig.~\ref{fig:c4h10-14-quantum-projection}) and a 28~V/\AA\ electric field (Fig.~\ref{fig:c4h10-28-quantum-projection}) show similar trends. The higher electric field increases the velocities of each atom consistently, corresponding to their increased charge states following ionization. This systematic increase reflects the more extensive ionization caused by the stronger field.

When comparing the classical simulations (14~V/\AA\ and 28~V/\AA) to their corresponding quantum simulations (Figs.~\ref{fig:c4h10-14-quantum-projection} and \ref{fig:c4h10-28-quantum-projection}), the simplified nature of the classical distributions becomes evident. The quantum simulations yield significantly broader velocity spreads centered around each atom, consistent with previous observations for C\textsubscript{2}H\textsubscript{2}. This broader distribution demonstrates the role of quantum effects in producing a more diverse set of velocities. Additionally, the quantum Newton plots consistently predict lower velocities for all atoms, aligning with earlier findings for C\textsubscript{2}H\textsubscript{2}. 

These differences persist across both electric field strengths (14~V/\AA\ and 28~V/\AA), indicating that quantum effects remain influential even in cases of stronger ionization that remove more valence electrons.

\subsection{Isoxazole (C\textsubscript{3}H\textsubscript{3}NO) Distributions}

\begin{figure*}[ht!]
    \centering
    \begin{subfigure}[b]{0.49\textwidth}
        \centering
        \includegraphics[width=\textwidth]{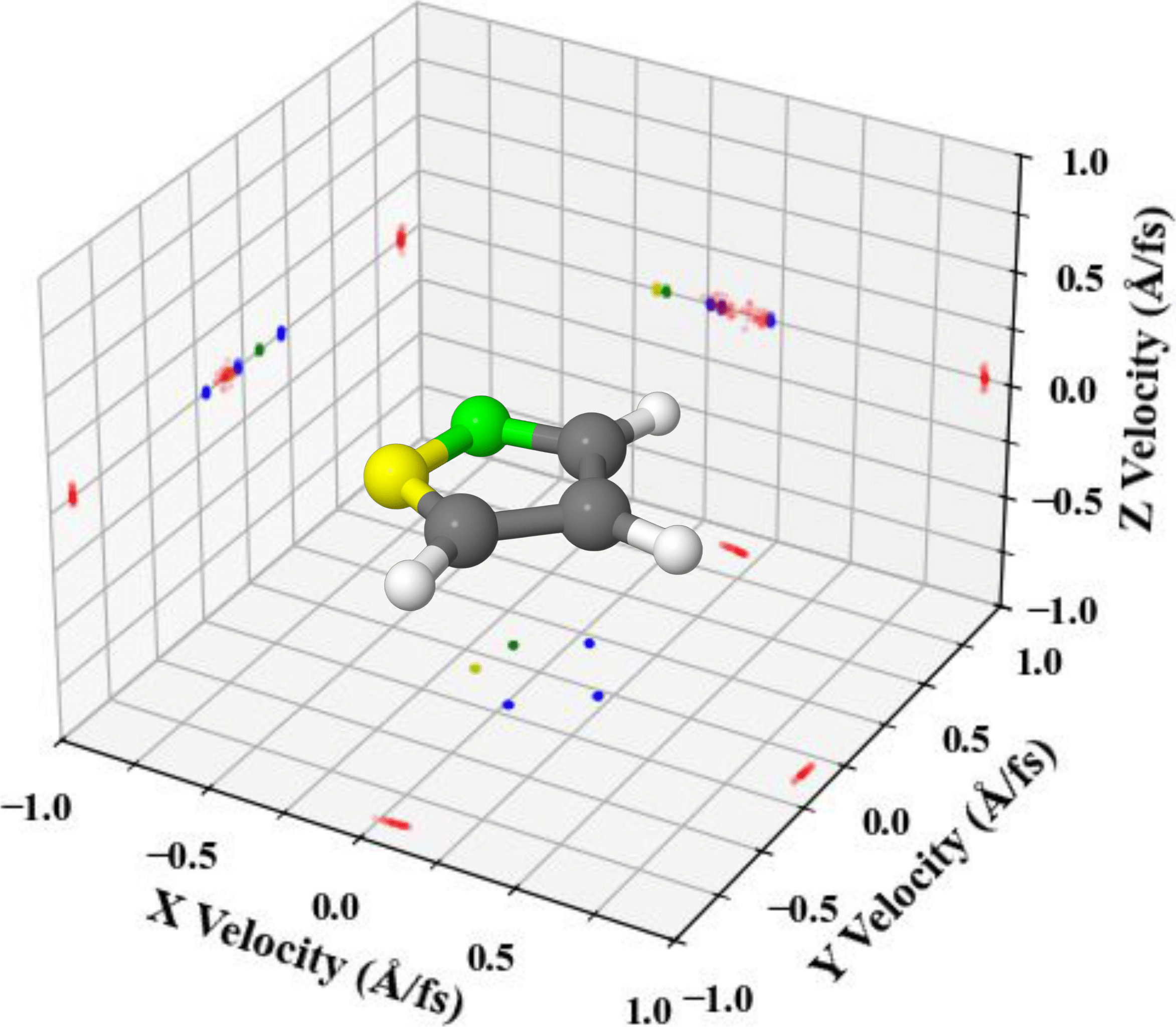}
        \caption{Classical method}
        \label{fig:isoxazole-classical-projection}
    \end{subfigure}
    \hfill
    \begin{subfigure}[b]{0.49\textwidth}
        \centering
        \includegraphics[width=\textwidth]{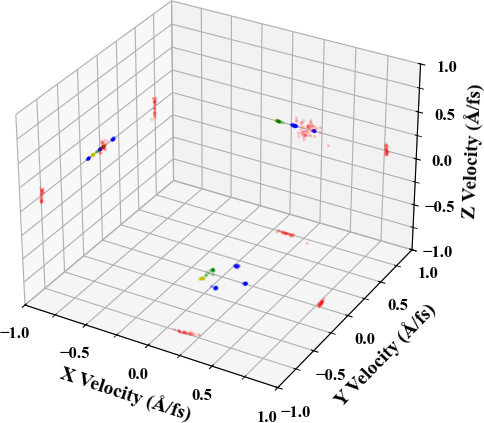}
        \caption{Quantum method}
        \label{fig:isoxazole-quantum-projection}
    \end{subfigure}
    
    \caption{3D projection plots of the final velocity of each carbon (blue), hydrogen (red), oxygen (green), and nitrogen (yellow) atom resulting from Coulomb explosion of isoxazole.}
    \label{fig:isoxazole-newton-plots-projection}
\end{figure*}

The final molecule investigated in this study is isoxazole. The choice of isoxazole as the target molecule was motivated by previous by the aforementioned recent study~\cite{PhysRevLett.132.123201}. The results from 20 Coulomb explosion simulations using each of the classical and quantum methods are presented in Fig.~\ref{fig:isoxazole-newton-plots-projection}. In these simulations, the three carbon atoms were assigned charges of $1.4^{+}$, $1.5^{+}$, and $1.5^{+}$, the three hydrogen atoms were each assigned a charge of $1^{+}$, and the oxygen and nitrogen atoms were each assigned a charge of $1.1^{+}$. The CAP was set at -12 to 12 \AA\ on the x-axis and -11 to 11 \AA\ on both the y- and z-axis.

The differences in velocity distributions across the methods shown in Fig.~\ref{fig:isoxazole-newton-plots-projection} are consistent with trends observed in other molecules: the angular distribution broadens, the diversity of simulation outcomes increases, and the velocities of all atoms decrease as quantum effects are introduced. Despite these differences, the overall structure of isoxazole remains distinguishable across the quantum method, as demonstrated by the overlap of the isoxazole ground state in Fig.~\ref{fig:isoxazole-classical-projection}. This aligns with experimental findings, which show that even in the presence of quantum effects, the molecular structure of isoxazole can still be reconstructed~\cite{PhysRevLett.132.123201}.

\section{Summary}
Quantum effects in Coulomb explosions were examined through a comparative analysis of three simulation methods: classical, semi-classical, and quantum. These effects arise from the explicit modeling of electrons in real space, their interactions with ions, and the ionization process. By treating electron densities explicitly, rather than as point charges localized at atomic nuclei as in classical methods, we gain valuable insights into their influence on the system and the discrepancies that emerge when Coulomb explosions are approximated using a classical model. Across all three molecules studied and various electric fields, quantum effects consistently lead to broader angular distributions, more diverse trajectory outcomes, and lower kinetic energies for all atoms as opposed to classical methods where the quantum effects are omitted.

These effects have the greatest influence in the early stages of the simulation when atoms are in close proximity and their charge states are evolving due to ionization. Once the atoms become sufficiently separated and reach their final charge states, quantum effects diminish, and the system behavior increasingly approximates classical dynamics.

The decrease in kinetic energy for all ions in the quantum model is primarily due to the presence of electrons and the gradual increase in ion charge during ionization. The electron densities act as points of attraction for ejected ions, thereby reducing their velocity. Furthermore, the laser causes ions to lose their electrons progressively, increasing their charge while the distance between ions also increases. Since the quantum model does not assume maximum post-ionization charge from the equilibrium geometry, the ions do not repel as strongly initially, further reducing their kinetic energies.

Experimental observations show that the momentum distribution of ions is significantly broader than what is predicted by classical simulations\cite{PhysRevLett.132.123201}. This discrepancy is attributed to the quantum effects observed in the present study. During and shortly after the laser pulse, electron dynamics play a pivotal role in influencing the motion of the fragmented ions. Initially, hydrogen atoms are ionized, and protons are ejected at the peak of the pulse. This is followed by carbon-carbon bond breaking, which occurs with a time delay. Furthermore, the oscillating electron densities during ionization provide additional points of attraction, introducing further variability into the trajectory outcomes and leading to broader angular distributions of the atoms.

Classical methods fail to accurately capture critical features such as the angular distribution of ion velocities, the final kinetic energy distributions of the ions, and the diversity of kinetic energy outcomes. In addition, these methods do not align as closely with experimental data from Coulomb explosion imaging experiments, where Newton plots are generated to reconstruct the molecular structure\cite{PhysRevLett.132.123201}. These limitations underscore the necessity of incorporating quantum effects to gain a more complete and accurate understanding of Coulomb explosion dynamics.

\begin{acknowledgments} 
This work has been supported by the National Science Foundation (NSF) under Grant No. DMR-2217759.

This work used ACES at TAMU through allocation PHYS240167 from the Advanced Cyberinfrastructure Coordination Ecosystem: Services \& Support (ACCESS) program, which is supported by National Science Foundation grants \#2138259, \#2138286, \#2138307, \#2137603, and \#2138296~\cite{aces}.

\end{acknowledgments}
%
%
%

\end{document}